\documentclass{ws-ijmpa}
\usepackage[super,compress]{cite}
\usepackage{verbatim}
\usepackage{graphicx}
\begin{document}
\markboth{RD52}{Fiber and Crystal Dual Readout calorimeters}

\catchline{}{}{}{}{}
%
%%%%%%%%%%%%%%%%%%%%%%%%%%%%%%%%%%%%%%%%%%%%%%%%%%%%%%%%%%%%%%%%%%%%

\title{Fiber and Crystals Dual Readout calorimeters}

\author{Michele Cascella}
\address{University College London, Gower Street, London WC1E 6BT, United Kingdom\\
m.cascella@cern.ch}

\author{Silvia Franchino}
\address{Kirchhoff-Institut for Physics, Heidelberg University, Germany\\
silvia.franchino@cern.ch}

\author{Sehwook Lee}
\address{Department of Physics, Kyungpook National University, 80 Daehak-ro, Buk-gu\\
Daegu, 41566, Republic of Korea\\
seh.wook.lee@cern.ch}

\author{on behalf of the RD52 collaboration.}

\maketitle

%\begin{history}
%\received{Day Month Year}
%\revised{Day Month Year}
%\end{history}

\begin{abstract}
The RD52 (DREAM) collaboration is performing R\&D on dual readout
calorimetry techniques with the aim of improving hadronic energy resolution
for future high energy physics experiments. The simultaneous detection
of Cherenkov and scintillation light enables us to measure the electromagnetic
fraction of hadron shower event-by-event. As a result, we could eliminate
the main fluctuation which prevented from achieving precision energy
measurement for hadrons. 

We have tested the performance of the lead and copper fiber
prototypes calorimeters with various energies of electromagnetic particles and
hadrons. During the beam test, we investigated the
energy resolutions for electrons and pions as well as the identification
of those particles in a longitudinally unsegmented calorimeter. 

Measurements were also performed on pure and doped PbWO$_{4}$ crystals,
as well as BGO and BSO, with the aim of realising a crystal based dual 
readout detector. We will describe our results, focusing on
the more promising properties of homogeneous media for the 
technique. Guidelines for additional developments on crystals will
be also given. 

Finally we discuss the construction techniques that we have used to
assemble our prototypes and give an overview of the ones that could
be industrialized for the construction of a full hermetic calorimeter.

\end{abstract}

%\ccode{PACS numbers:}

%\tableofcontents

\section{Introduction}

The role of calorimeter in modern high energy physics experiments
has been identifying particles as well as measuring their energy and
momentum. Such experiments, the calorimeters have provided
the key information on the measured energy and momentum of particles
identified as the electromagnetic particles and hadrons by itself,
and have become the heart of particle physics experiments. From this
fact, we can infer that the measurement of particle energy with excellent calorimeter energy resolution ends up with high quality particle
physics experiments and good physics results. After the Higgs discovery,
it is believed that a future lepton collider will focus on understanding
the Higgs mechanism, and study of the Higgs boson properties. 
To accomplish this, we will need a dedicated Higgs factory with excellent
detectors which can measure the energy of jets from hadronic decays
of W's and Z's in the same precision as the energy measurement of
electrons and gammas. \\
However, while the detection of electrons, photons and other particles
that develop electromagnetic showers can be performed with high precision,
the same is not true for hadrons.\\
This is primarily due to the fact that 
\begin{enumerate}
\item most calorimeters generate a larger signal per unit deposited
energy for the electromagnetic shower component ($e$) than for the
hadronic one ($h$); that is $e/h>1$ 
\item the fluctuations in the energy sharing between these two components
are large and non-Poissonian. 
\end{enumerate}
As a result, in typical instruments the hadronic response function
is non-Gaussian, the hadronic signals are non-linear.

Several approaches have been proposed to deal with this problem: 
\begin{description}
\item [{Compensating~calorimeters}] are designed to deliver equal response
to the $em$ and non-$em$ shower components: $e/h=1$. This can be
achieved by boosting the response to the hadronic component, for instance
in calorimeters with a hydrogenous active medium that is very sensitive
to the soft neutrons abundantly produced in hadronic shower development.
Such calorimeters require a precisely tuned sampling fraction. Since
this sampling fraction is typically small (2.3\% in lead/plastic-scintillator
calorimeters), the em energy resolution of such devices is in practice
limited to $15\%/\sqrt{E}$\cite{spacal}. 
\item [{Off-line~compensation}] is a technique applied in devices with
$e/h\ne1$, in which signals from different sections of the calorimeter
are re-weighted according to some scheme to reconstruct the shower
true energy. This is one of the most common approach, however, it
requires attention to avoid introducing non linearities in the response. 
\item [{Energy~Flow}] combines the calorimeter information
with measurements from a tracking system, to improve the performance
for jets. This method, usually deployed in combination with the previous
one, has been successfully applied in several modern collider experiments. 
\item [{Particle~Flow}] extends the previous technique by trying to  reconstruct each individual particle in a jet using
highly segmented calorimeters and sophisticated reconstruction algorithms.
\item[{Dual~Readout~Method}] that is the topic of this paper and will be fully described in the next section.
\end{description}

\subsection{The Dual REAdout Method}

Since the resolution is determined by fluctuations in the electromagnetic fraction of the shower ($f_{em}$), measurement of the $f_{em}$ value, event by event, is the key to improving the hadronic energy resolution of an intrinsically non-compensating calorimeter.
In our method, $f_{em}$ is measured by comparing the shower signals
produced in the form of Scintillation light ($S$) and Cherenkov light
($C$) in the same detector.

In dual-readout calorimetry each hadronic shower is measured in two
nearly independent ways. While the Scintillation channel is sensitive
to both components of the shower, the signals from the hadronic fraction
are strongly dominated by spallation hadrons produced in nuclear reactions,
these are usually not sufficiently relativistic to produce Cherenkov
light. The Cherenkov channel is, for all practical purposes, only
sensitive to the em shower component ($e/h\gg1$).

This can be realized in different ways: 
\begin{itemize}
\item In a fiber calorimeters, scintillating fibers and clear fibers can
used to measure the S and C channel separately. 
\item In a crystal calorimeter, both scintillation and Cherenkov light are
generated in the same optical volume, and the necessary separation
of the two kinds of light is accomplished using time structure, direction,
wavelength spectrum, and polarization. 
\end{itemize}
The $e$ part is calibrated to unit response, and the resulting average
response of the $h$ part is denoted by $\eta=h/e$, which is less
then unity for most calorimeters. If $E$ is the
hadronic energy (either single hadron or jet), and the responses expected
in the two channels are 
\[
S=E\left[f_{em}+(1-f_{em})\eta_{S}\right]
\]
\[
C=E\left[f_{em}+(1-f_{em})\eta_{C}\right]
\]

These equations can be inverted to measure both $E$ and $f_{em}$
event-by-event.

\begin{eqnarray}
E_{meas}=\frac{S-\chi C}{1-C}~{\rm {with}~\chi=\frac{1-\eta{}_{S}}{1-\eta_{C}}}\label{eq-dream}
\end{eqnarray}

Where $E_{meas}$ is the corrected energy which is a function of the
Scintillation and Cherenkov signals, and the factor $\chi$, which,
in turn, is determined by the physical characteristic of the calorimeter.

\section{The RD52 Fiber Calorimeter}

\begin{figure}
\centerline{\includegraphics[width=0.95\textwidth]{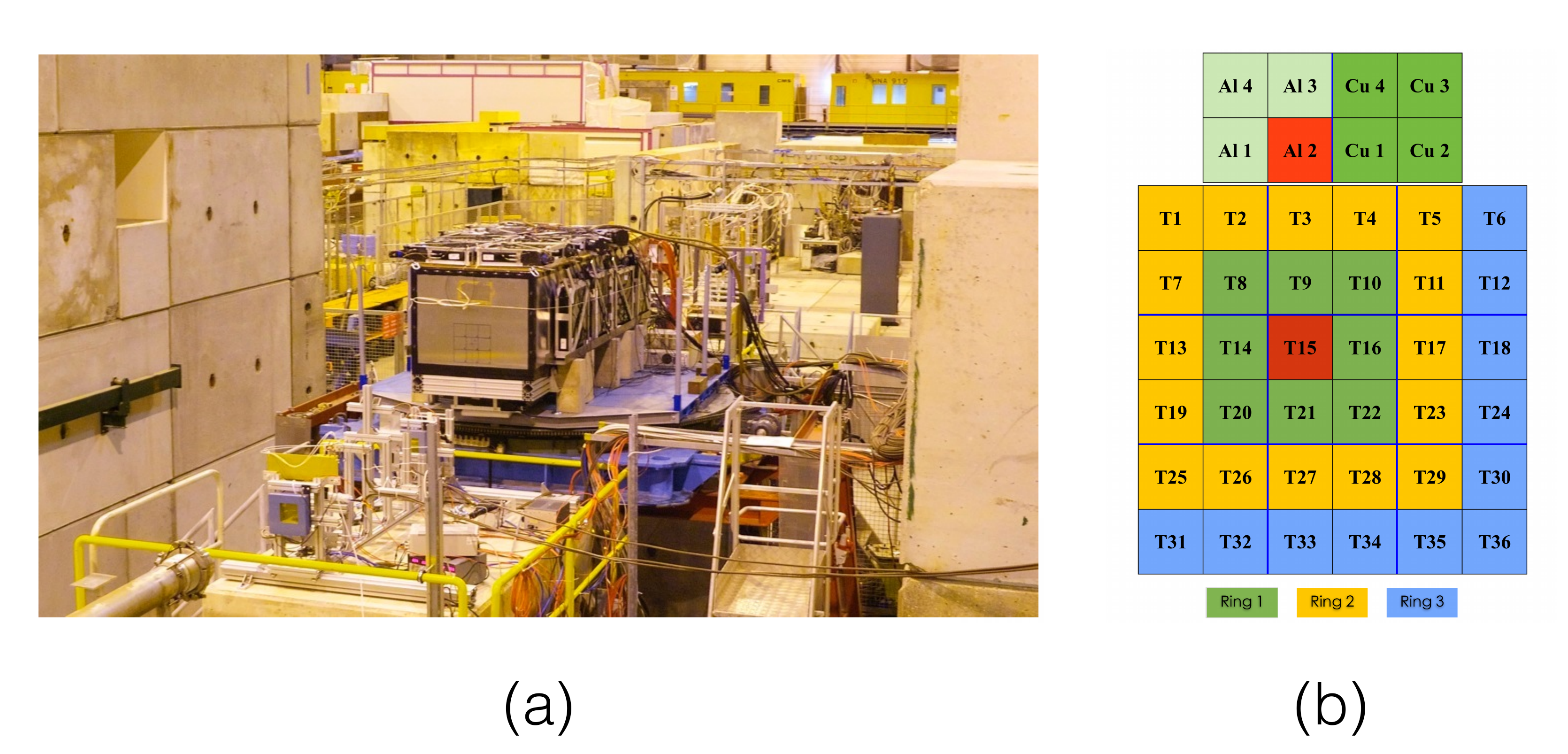}} \caption{The beam test setup in the H8C area of SPS at CERN in Nov. 2012 (a),
and the tower map of the Pb- and Cu-fiber calorimeters (b). The front end of the Cherenkov fibers in the left copper module have been aluminised on the front end to increase the light output. \label{fig:exp_setup}}
\end{figure}

The RD52 collaboration has undertaken an extensive experimental campaign to 
validate the dual readout method with fiber calorimeters. 
We have a semi permanent setup at H8 beam line in the CERN SPS North 
Area. The experimental setup for the Nov. 2012 beam test is shown in Fig.~\ref{fig:exp_setup} and all experimental results described in this paper have been obtained using a substantially unmodified set of detectors.

A number of auxiliary detectors are installed on the beam line:

\begin{description}
\item [{Drift~chambers}] (DC) the two wire chambers are used to constrain
the particle position and divergence to obtain a well $10\times10\,\text{mm}^{2}$
beam spot.
\item [{Preshower~detector}] (PSD) a scintillator placed right after a
5~mm thick Pb plate. Electrons start developing showers in this device,
while muons and hadrons typically produced a signal characteristic
for a minimum ionizing particle (mip) in the scintillator plate. 
\item [{Tail~catcher}] (TC) a $20\times20\,\text{cm}^{2}$ scintillation
counter placed right after the calorimeter ($10\lambda_{int}$), used
to identify escaping pions and muons.
\item [{Muon~counter}] (MuC)a $50\times50\,\text{cm}^{2}$ scintillator
placed right after the tail catcher behind an additional $8\lambda_{int}$
of iron.
\end{description}

Elimination of the hadron (electron) and muon contamination in the electron (hadron) beams is performed using these ancillary detectors.

The RD52 fiber calorimeters consisting of 9 Pb-fiber and 2 Cu-fiber
modules were housed by the rectangular box, which is shown in the
middle of Fig.~\ref{fig:exp_setup} (a).

Each module is 2.5~m long, corresponding to $\sim10\lambda_{int}$, and
its cross section is $92\times92$~mm$^{2}$. Each tower is a square of 
46$\times$46 mm$^{2}$ and has one Cherenkov and one scintillation channels; 
one module consists of four towers. Fig.~\ref{fig:exp_setup} (b) shows how the 9 Pb-fiber modules are arranged in a 3$\times$3 matrix and the relative position of the 2 Cu-fiber modules that are placed on top of the Pb matrix. Tower 15 (T15) is defined as the central tower of the matrix, surrounded by three rings. The results described here were obtained with beams aimed at T15 (for the Pb 
matrix) or Al2 (for the Cu-fiber calorimeter). 

The cross section of 3$\times$3 matrix is 27.6$\times$27.6
cm$^{2}$. It is small compared to the average lateral size of an hadronic shower, so shower leakage affects the performance of the RD52 calorimeters. To try and count for that, 20 leakage counters, made of 50$\times$50$\times$10 cm$^{3}$
of plastic, are installed around the RD52 fiber calorimeters. 

\begin{comment}
\begin{figure}
\centerline{\includegraphics[width=0.95\textwidth]{cal_structures}}
\caption{Basic structures of Pb-fiber (a) and Cu-fiber (b) modules\label{fig:cal_structure}}
\end{figure}

Fig.~\ref{fig:cal_structure} contains pictures which show the structures of Pb- (a) and Cu-fiber (b) modules. In Fig.~\ref{fig:cal_structure} (a), scintillation and Cherenkov fibers were put in alternating rows of grooves which were made on the both sides of a Pb plate. Differently from the Pb-fiber structure, Fig.~\ref{fig:cal_structure} (b) shows that copper was eliminated to make grooves on one side whose diameter of 1.0 mm. Embedding scintillation and Cherenkov fibers alternately forms the main pattern of the basic Cu-fiber structure. 
\end{comment}

\subsection{Electromagnetic performance}

Each tower of the RD52 calorimeter contains an equal number of scintillating and clear fibers (where only Cherenkov light is produced). Both channels of every individual tower are calibrated using the response to 20 GeV
electrons. 

\begin{figure}
\centerline{\includegraphics[width=0.75\textwidth]{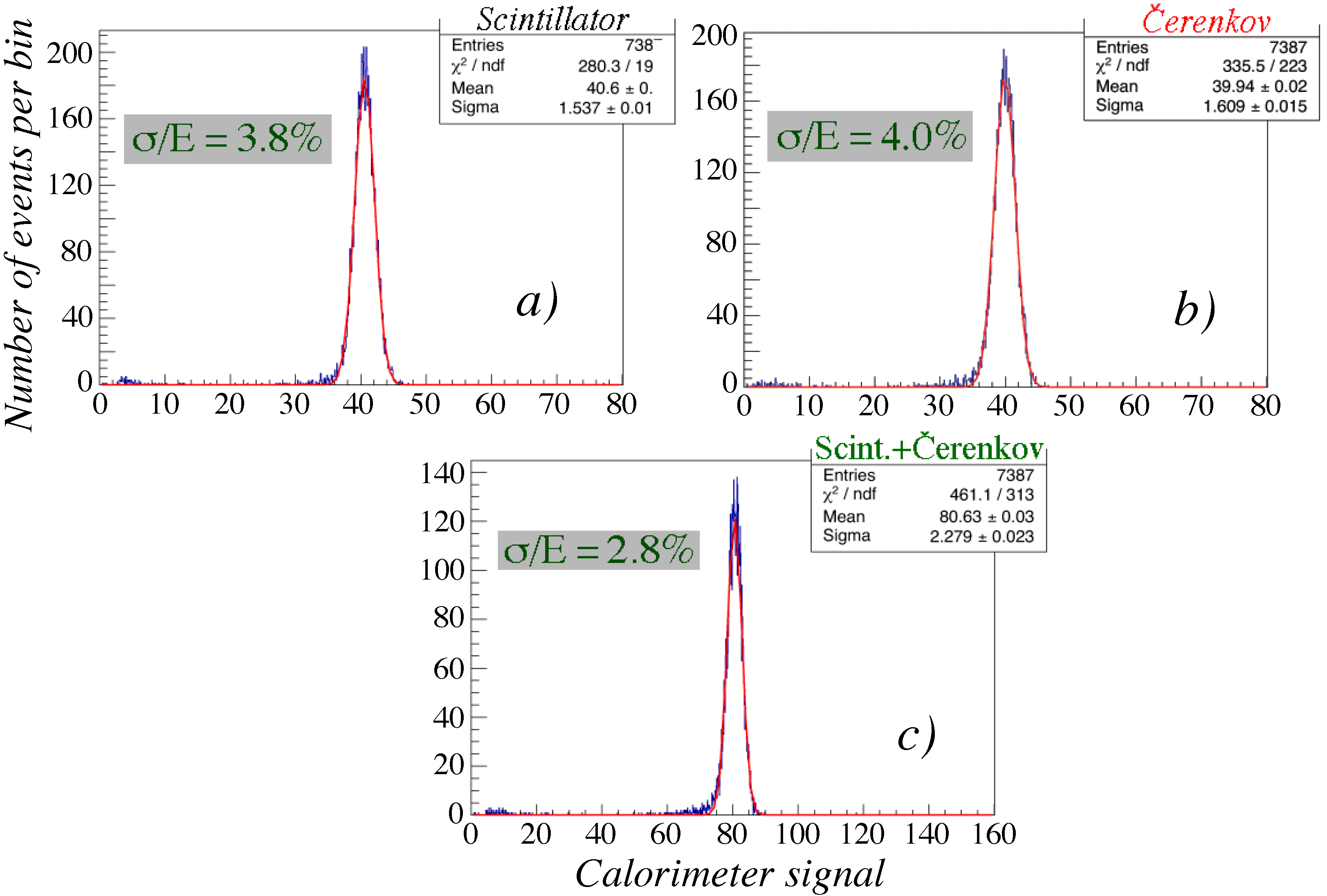}} \caption{Signal distributions of the scintillation (a), Cherenkov (b) channels,
and the combination of two types of signals (c) for 40 GeV electrons.\label{fig:40GeV_e}}
\end{figure}

Fig.~\ref{fig:40GeV_e} shows the response of the aluminised Cu-fiber 
calorimeter to 40 GeV electrons. In both scintillation and Cherenkov channels 
the electron energy is measured with a resolution of $\sigma_{E}/E$ of 3.8\% and 4.0\% respectively. The signal distributions of the scintillation and Cherenkov channel is plotted in Fig.~\ref{fig:40GeV_e}a and b. 

Since the same shower is sampled by the Cherenkov and scintillation fibers 
at the same time the two signals can be simply combined giving an improved 
energy resolution of $\sigma_{E}/E=2.8\%$.

%\begin{figure}
%\centerline{\includegraphics[width=0.45\textwidth]{figures/Figure_13}} \caption{The %energy resolution of the Cu-fiber calorimeter for electrons as
%a function of 1/$\sqrt{E}$.\label{fig:em_res}}
%\end{figure}

\begin{figure}
\centerline{\includegraphics[width=0.48\textwidth]{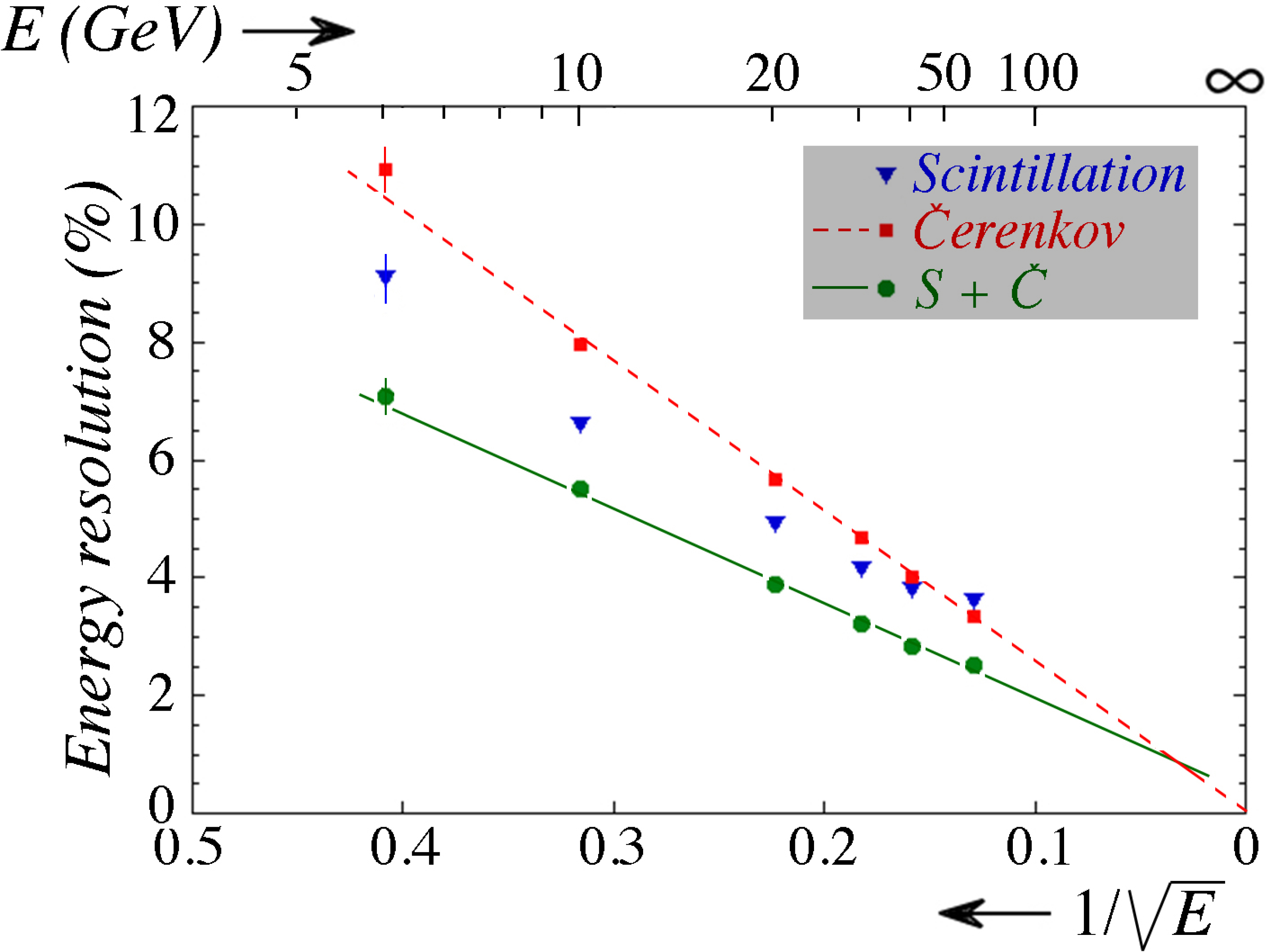}\includegraphics[width=0.48\textwidth]{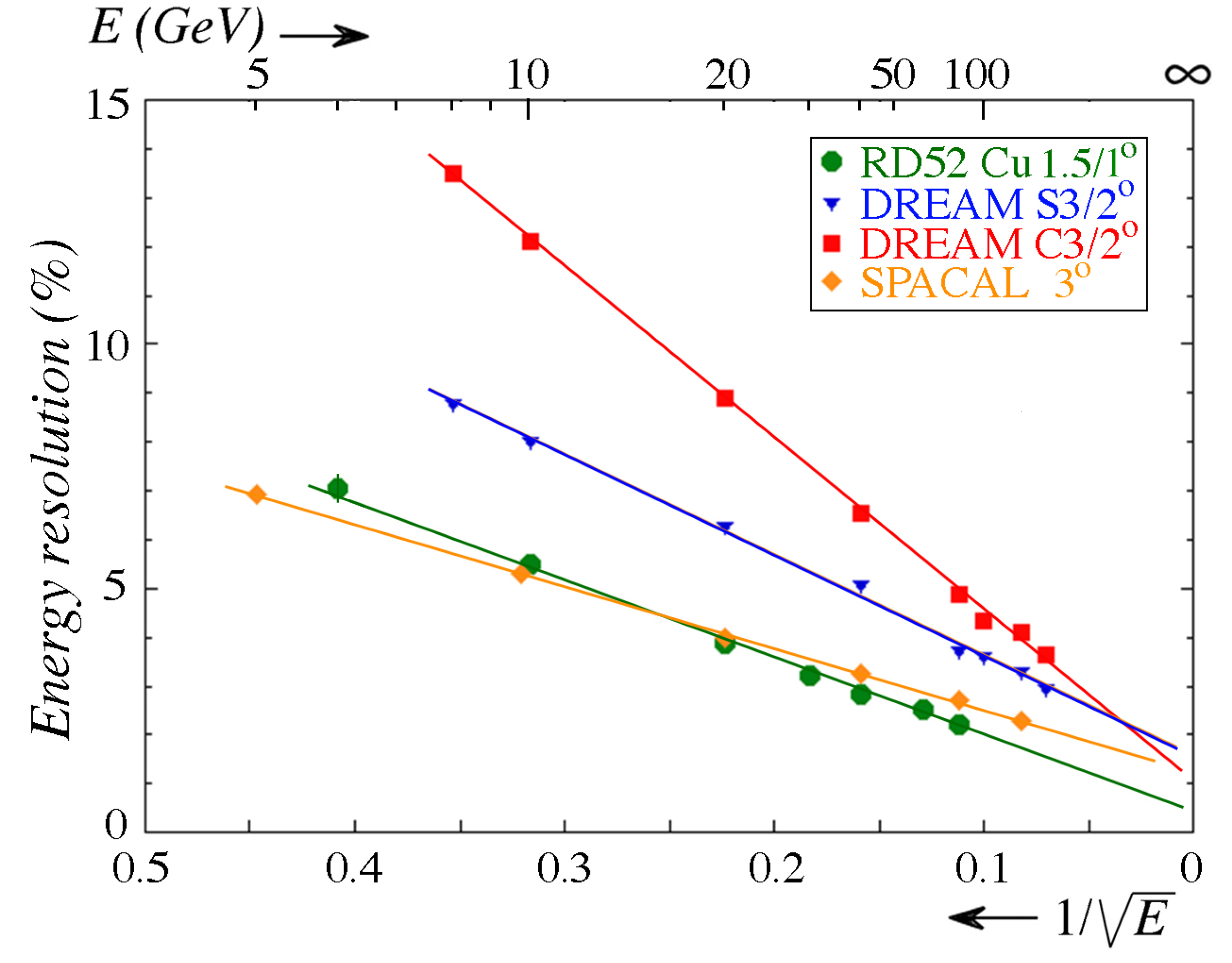}} \caption{The energy resolution of the Cu-fiber calorimeter for electrons as a function of 1/$\sqrt{E}$ (right).\label{fig:em_res} Comparisons of the em energy resolutions of the prototype DREAM, RD52 calorimeters and SPACAL (left). \label{fig:res_comp}}
\end{figure}

To investigate the electromagnetic performance, the Cu-fiber module was
exposed to various electron energies. Fig.~\ref{fig:em_res} shows
the energy resolution as a function of $1/\sqrt{E}$. The blue inverted
triangles and the red squares are the energy resolutions for the scintillation
and Cherenkov channels, respectively. The Cherenkov channel is well
described by the straight line, and this suggests a very small
 constant term. The scintillation channel, on the other hand, deviates from the linear behaviour by 2 to 3\%.
This deviation comes from the response difference between
particle hitting the absorber first or the scintillation fibers first. 

Combining the two signals, the energy resolution is clearly
improved as the green circles in Fig.~\ref{fig:em_res} show. The
green straight line fit results in a resolution $${\frac{\sigma_{E}}{E}} = {\frac{13.9\%}{\sqrt{E}}}$$ with constant term of less than 1\%.

%\begin{figure}
%\centerline{\includegraphics[width=0.75\textwidth]{figures/Figure_22}} %\caption{Comparisons of the em energy resolutions of the prototype DREAM, RD52
%calorimeters and SPACAL. \label{fig:res_comp}}
%\end{figure}

It is useful to compare the electromagnetic performances of the RD52 calorimeter with that of other fiber calorimeters such as SPACAL\cite{spacal} and the old DREAM
(the first prototype dual-readout calorimeter)\cite{dream}.
Fig.~\ref{fig:res_comp} shows that the RD52 calorimeter has
better energy resolution than DREAM (thanks to the improved sampling fraction). Moreover the energy resolution of our copper calorimeter comparable to SPACAL and better than that
for energies larger than 20 GeV.

\subsection{Hadronic performance}

\begin{figure}
\centerline{\includegraphics[width=0.95\textwidth]{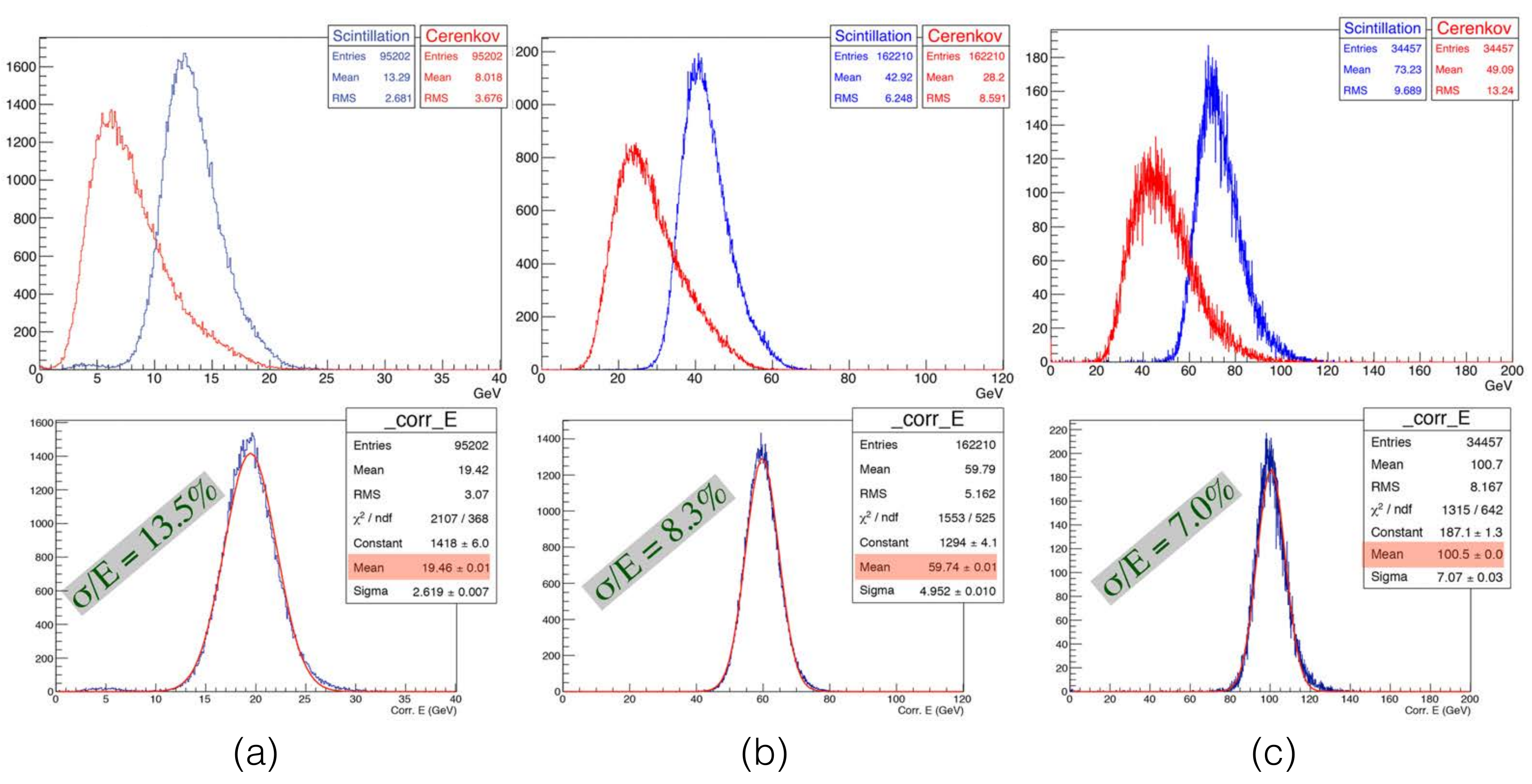}} \caption{The 
response functions of the scintillation and Cherenkov channels
to pions before (upper) and after (lower) the dual-readout method
correction for 20 GeV, 60 GeV, and 100 GeV\label{fig:hadron_res}}
\end{figure}

Fig.~\ref{fig:hadron_res} (a), (b), and (c) shows the Pb-fiber
calorimeter responses to of 20 GeV, 60 GeV, and 100 GeV pions, respectively.
The top three plots show the scintillation (blue) and Cherenkov (red)
signal distributions for those three energies. These are typical
non-compensating calorimeter response to hadrons. Their response
functions are non-Gaussian and the average energies measured are lower than
the beam energies. The energy resolutions in the scintillation channel are 20\%, 15\%, and 13\% respectively.

However, we can use the dual-readout method described in Eq.~\ref{eq-dream}
to reconstructs the corrected beam energies with Gaussian response functions,
and significantly improve the energy resolution. After the dual-readout correction 
the resolutions become 13.5\%, 8.3\%, and 7.0\% respectively.

\begin{figure}
\centerline{\includegraphics[width=0.75\textwidth]{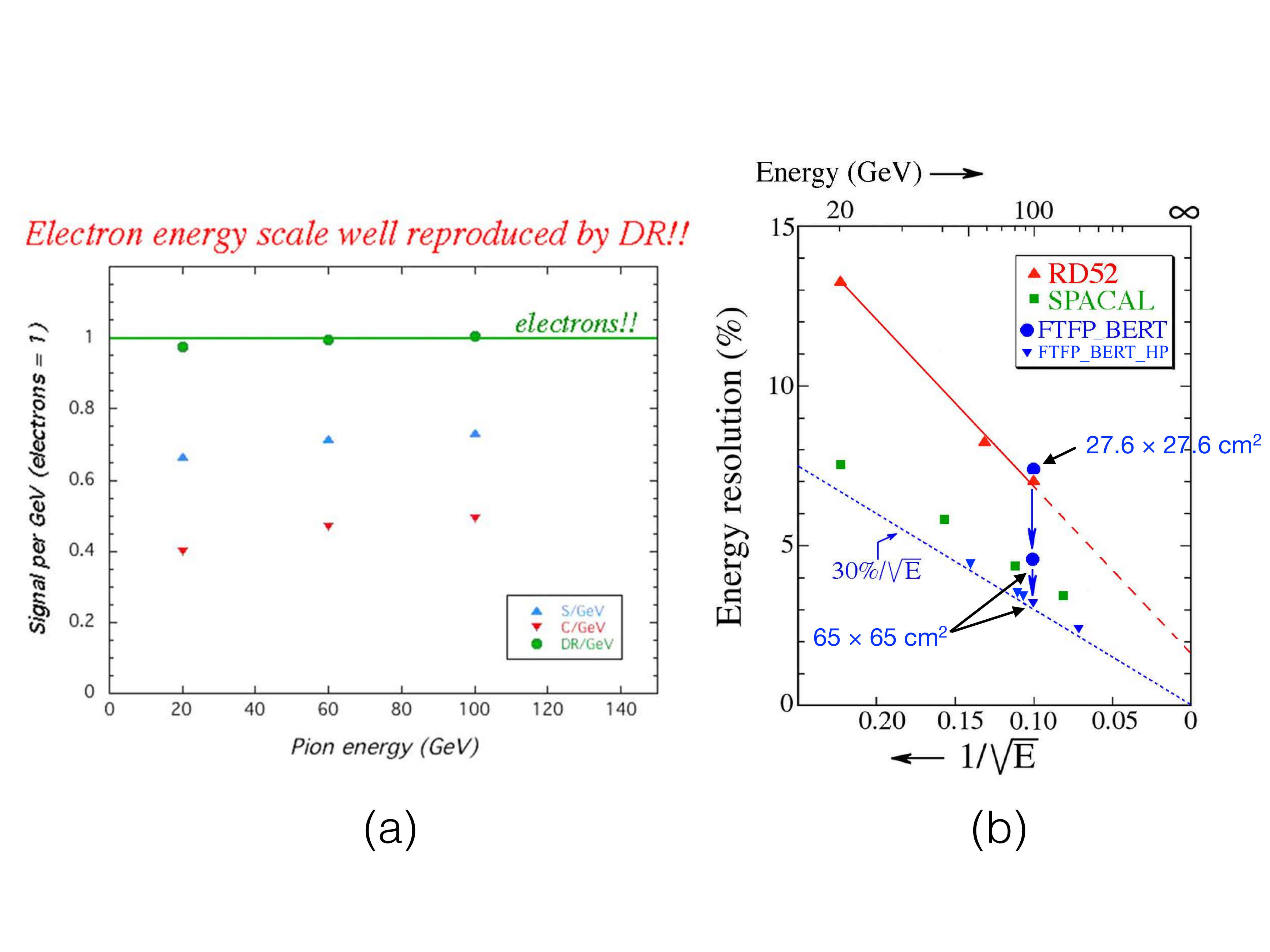}}
\caption{The RD52 calorimeter responses to pions for the scintillation (blue 
triangles) and Cherenkov (red inverted triangles) channels, and after the 
dual-redout correction (green circles) (a). Comparison
of the hadron energy resolutions of the RD52 expereimental data (red triangles), the RD52 simulation (blue circles for FTFP\_BERT and blue inverted triangles for FTFP\_BERT\_HP), and SPACAL (green squares)
(b). \label{fig:dr_res}}
\end{figure}

Fig.~\ref{fig:dr_res} (a) shows the calorimeter responses as a function of pion 
energy for the scintillation, Cherenkov, and combined channels. The responses of 
the scintillation and Cherenkov channels are, on average, 70\% and 45\%, of that
of electrons of the same energy. But the dual-readout method correction restores the electron energy scale and the linearity of the response.

As we have already mentioned the RD52 calorimeter lateral size is smaller than the average lateral size of the typical hadronic shower and its resolution is negatively affected by shower leakage. To see how a bigger
module can improve the hadronic performance, we simulated the performance of a 65$\times$65 cm$^{2}$ RD52 fiber matrix with GEANT4~\cite{geant4} for 50, 80, 90, 100 and 200 GeV pions. We used two physics lists such as FTFP\_BERT
and FTFP\_BERT\_HP\cite{phy_list}, the second one being much slower but 
with a high precision neutron model used for neutrons below 20~MeV.

In Fig.~\ref{fig:dr_res} (b), we compare the results obtained with the RD52 calorimeter and SPACAL, with the simulation. 
%SPACAL has better hadron energy resolution than RD52. 
SPACAL with a stochastic term of $30\%/\sqrt{E}$ still holds
the world record for the hadron energy resolution. It was a relatively
larger detector than the RD52 calorimeter, so that the leakage fluctuation
has a negligible contribution to the the hadronic energy resolution. 

First we simulated a dual readout Pb-fibre calorimeter the same size of the real RD52 Pb matrix (27.6$\times$27.6~cm$^{2}$) using the FTFP\_BERT physics list. The performance of this detector with 100~GeV pions is shown with a blue circle in Fig.~\ref{fig:dr_res} (b) and is consistent with the experimental results. 

Following this result, we investigated the hadron performance of a larger version of the RD52 calorimeter (65$\times$65~cm$^{2}$, equivalent to a 4x4 matrix)  using the FTFP\_BERT\_HP for 50, 80, 90, 100 and 200 GeV pions (see Fig.~\ref{fig:dr_res} b). These simulations indicate that a dual readout calorimeter large enough (and in a collider experiment it certainly would be) to have negligible  lateral shower leakage  would have an hadronic resolution of about $30\%/\sqrt{E}$ and a very small constant term. 

Our simulation predicts that the dual-readout method
calorimeter can achieve the high-quality hadron and jet energy measurement
with a much better electromagnetic resolution than that of a compensating calorimeter.

\begin{comment}
Using the simulation we estimate
how precisely a larger RD52 calorimeter measures hadron energy. First
off, we constructed a RD52 calorimeter which was 2.5 m long and  of cross section. This is the exactly same size of the real
RD52 calorimeter. With this detector, simulated events with GEANT4
were produced and the energy resolution was measured for 100 GeV pions.
This result is the blue circle in Fig.~\ref{fig:dr_res} (b), which
is consistent with the experimental result of the RD52 calorimeter
denoted with the . Based on this result, the hadron performance
for 65$\times$65 cm$^{2}$ of the calorimeter was investigated using
GEANT4 for 50, 80, 90, 100 and 200 GeV pions. GEANT4 simulation suggests
that the RD52 calorimeter have $30\%/\sqrt{E}$ and very small constant
term for pions. The simulation predicts that the dual-readout method
calorimeter can achieve the high-quality hadron and jet energy measurement
without the limitations of compensating calorimeters.
\end{comment}

\subsection{Particle identification}

Identification of isolated electrons, pions and muons is of particular
importance in particle colliders, for the study of the decay of Higgs
bosons into pairs of $\tau$ leptons. In the following we discuss
the possibly of doing particle separation with the RD52 Pb-fibre calorimeter.
Such an ability would be a great asset for an experiment at a future
Higgs factory.

%\subsubsection{Data analysis}

The particle beams in the H8 beam line typically contain a mixture of electrons, pions and muons. Using the auxiliary detectors described above we can select the following samples:
\begin{description}
\item [{Electrons}] are identified as particles that produced a signal
larger than that of 2 minimum ionizing particles (mips) in the PSD.
We also require a signal compatible with electronic noise in the TC
and MuC. The total scintillation signal in the calorimeter should
be larger than 15 GeV for the 20 GeV beam and larger than 50 GeV for
the 60 GeV beam.
\item [{Pions}] are particles that produced a signal compatible with a
mip traversing in the PSD, and a signal compatible with noise in the
MuC. The total scintillation signal in the calorimeter should be larger
than 7 GeV.
\end{description}

%\subsubsection{Results}

We have developed several techniques techniques to identify the nature
of a particle using our Pb-fiber calorimeter. 

\begin{figure}
\begin{centering}
\includegraphics[width=0.45\textwidth]{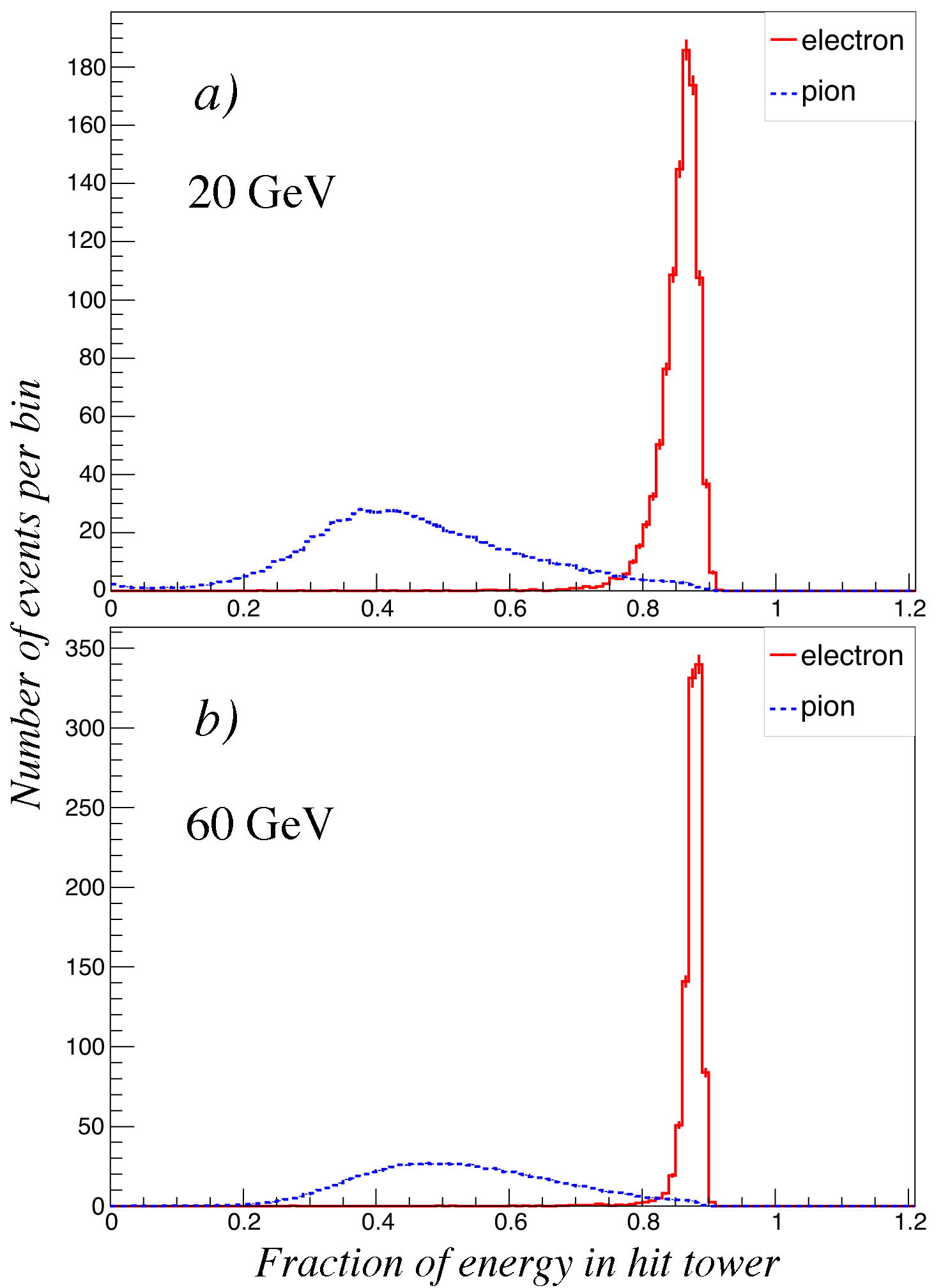}\includegraphics[width=0.44\textwidth]{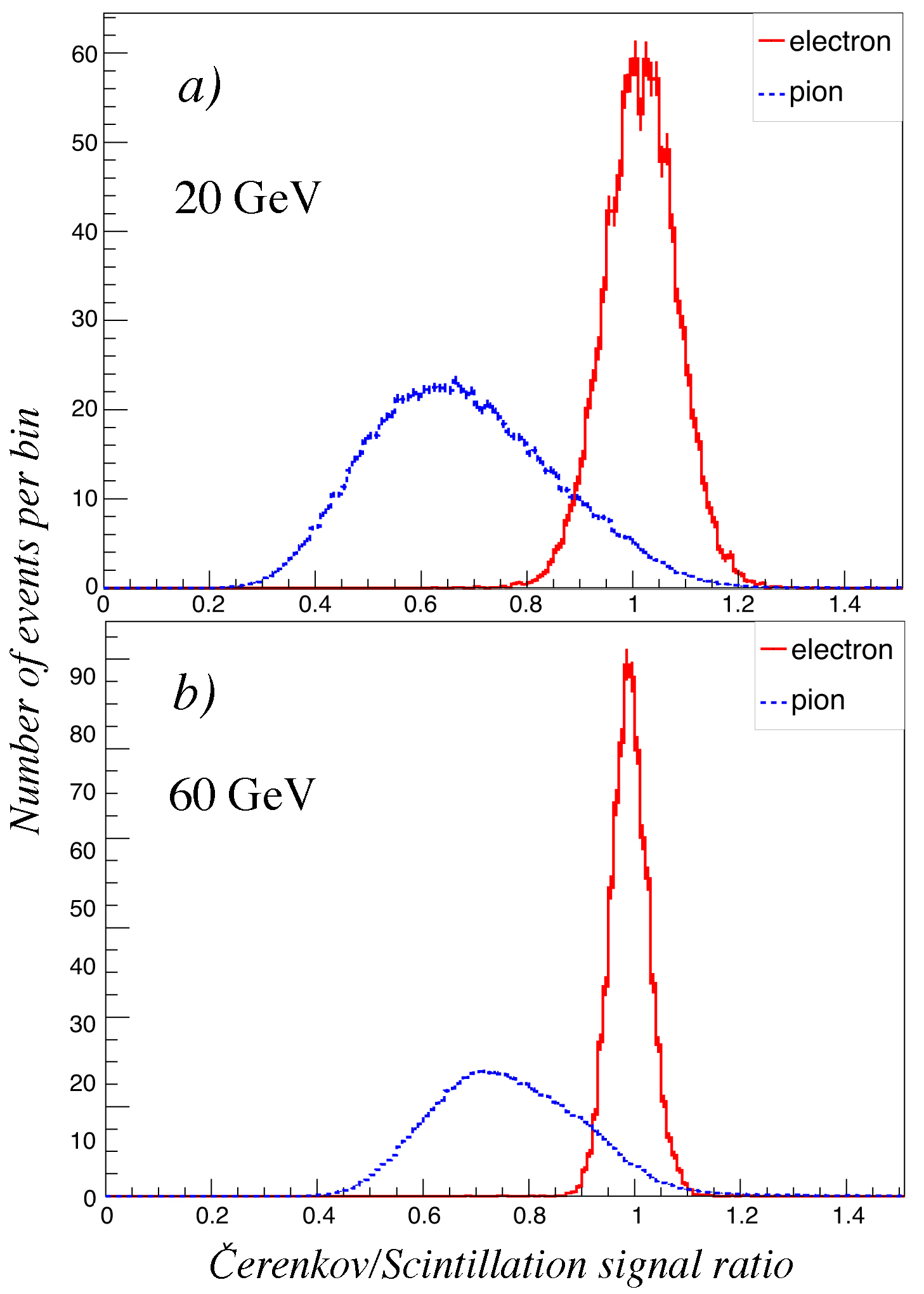}
\par\end{centering}
\caption{\label{fig:Distribution-of-e-fraction}\label{fig:C-S-ratio}Distribution
of the energy fraction deposited in the hit tower by electrons and
pions of 20 GeV (top left) and 60 GeV (bottom left). Distribution
of the C/S signal ratio in the hit tower for 20 GeV (top right) and
60 GeV (bottom right) electrons and pions.}
\end{figure}

The first separation method uses the lateral shower size to distinguish
between em and hadronic showers. The calorimeter towers have a lateral
size of 1.6 Moliere radii, or 0.2 nuclear interaction lengths. Our
measurements show that electrons hitting the center of a tower deposit
typically 85\% of their energy in that tower, while hadrons typically
deposit only 40\textendash 50\%; this is shown in Fig.~\ref{fig:Distribution-of-e-fraction}.
\begin{comment}
The average energy deposited by pions in the hit tower clearly increases
with the energy of the beam particles as the average em shower fraction
increases with the hadron energy.
\end{comment}

A unique aspect of a dual readout calorimeter is the fact that two
types of signals are produced: scintillation (S) signals and Cherenkov
(C). The ratio of the two types of signals, $C/S$, is typically around
1 for electron showers, while it is smaller than 1 for hadron showers.
Fig.~\ref{fig:C-S-ratio} shows the distribution of the C/S signal
ratio for electrons and pions, at energies of 20 and 60 GeV. %
\begin{comment}
The distribution for the pions shifts to larger values, from an average
of 0.6 to 0.7, and the reasons are the same as the ones given for
the lateral shower profile
\end{comment}
The width of the electron distribution shrinks because of the reduced
event-to-event fluctuations, while the average value of the pion distribution
increases because of the increased em shower fraction. 

\begin{figure}
\begin{centering}
\includegraphics[width=0.75\textwidth]{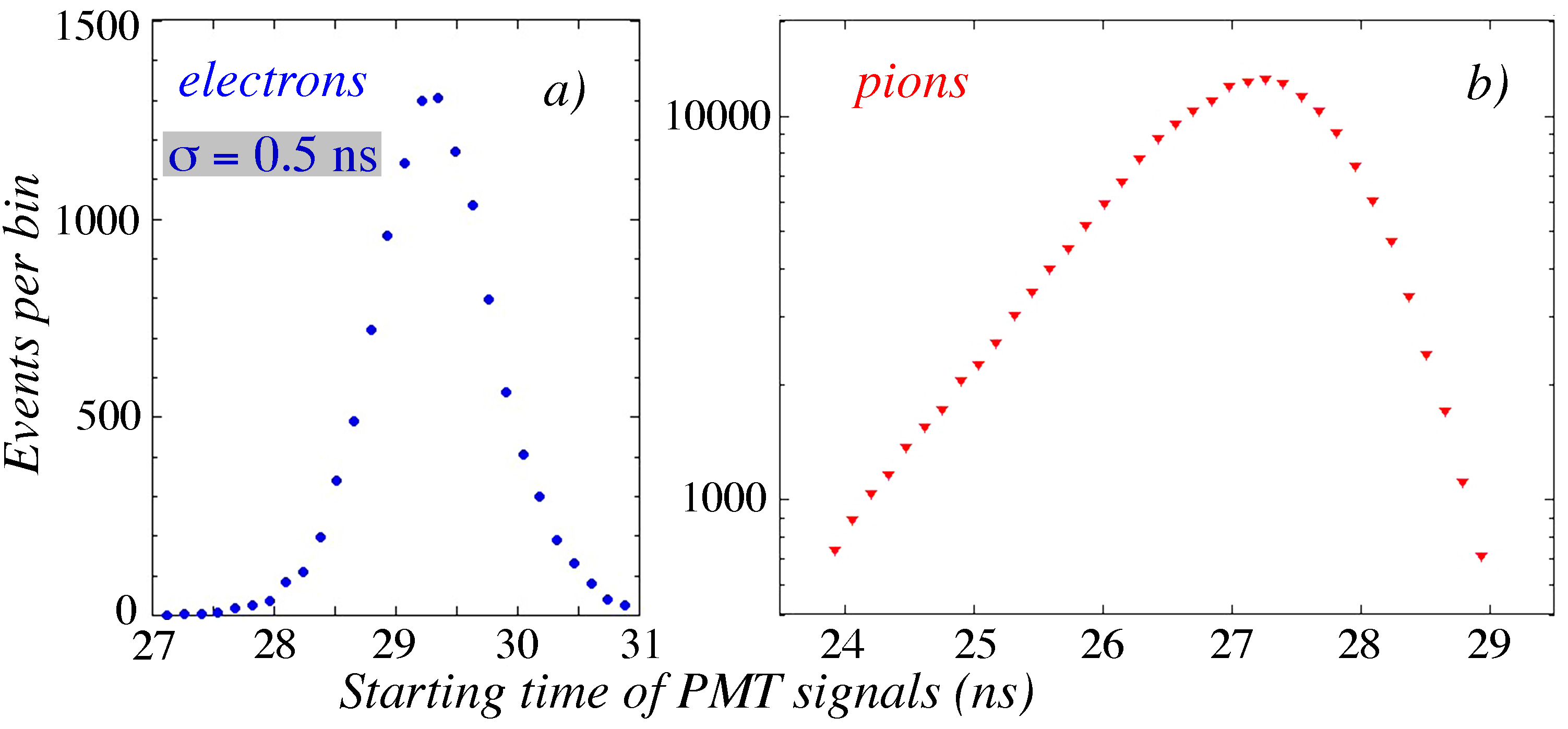}
\par\end{centering}
\caption{\label{fig:starting-time}The measured distribution of the starting
time of the calorimeter's scintillation signals with respect to the
trigger signal produced by 60 GeV electrons (a) and 60 GeV pions (b). }
\end{figure}

Measurements of the average depth at which the energy is deposited
inside the calorimeter provide a powerful tool for particle identification.
In longitudinally unsegmented calorimeters, this depth can be measured
using the fact that the light in the optical fibers travels at a lower
speed ($\sim17\,\text{cm/ns}$ for polystyrene based fibers) than
the particles that generate this light (close to $c\sim30\,\text{cm/ns}$).
As a consequence the the deeper inside the calorimeter the light is
produced, the earlier the PMT signal. The depth of the light production
can be determined for individual events can be determined with a precision
of $\sim20\,\text{cm}$. Fig.~\ref{fig:starting-time} shows the
measured TDC distribution for 60 GeV pions and electrons; pions peak
1.5~ns earlier than electrons. The effect of the PSD on this measurements
is estimated to be smaller than $0.02\,\text{ns}$.

\begin{figure}
\begin{centering}
\includegraphics[width=0.75\textwidth]{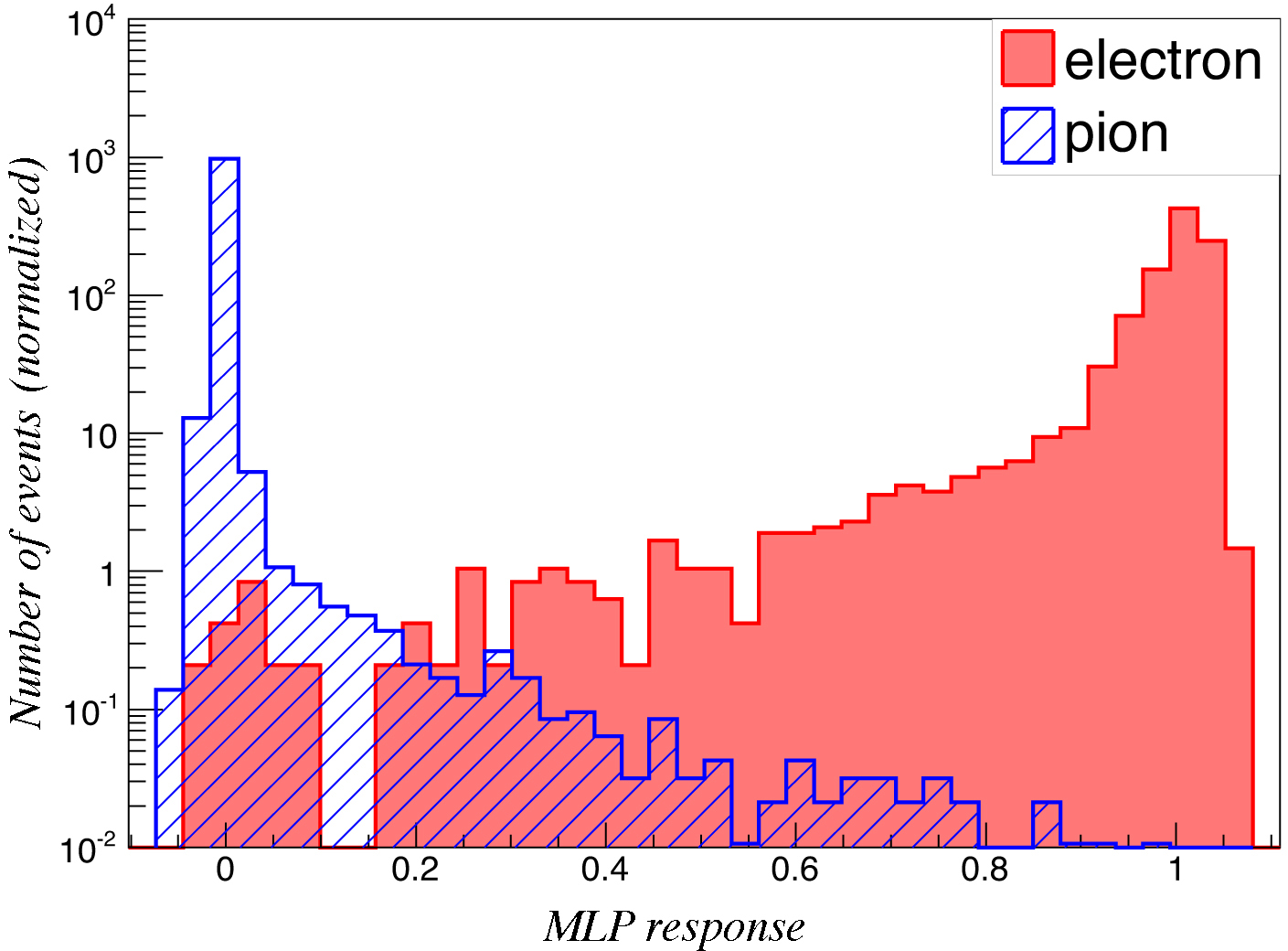}
\par\end{centering}
\caption{\label{fig:MVA}Results from the multivariate analysis of the electron/pion
separability at 60 GeV, using the lateral shower profile, the C/S
signal ratio and the starting time of the PMT signals together to
separate electrons from pions. }
\end{figure}

A high degree of $e/\pi$ separation can be achieved by combining
different methods: using a multi-layer perception can identify 99.8\%
of all electrons with 0.2\% pion contamination (see Fig.~\ref{fig:MVA}). 

More information on these studies, including the treatment of another
method based on the PMT pulse width, can be found in \cite{partcle_ID}.

\section{Construction of the RD52 fiber prototypes}

The tested calorimeter prototypes were built in 2012 inside the INFN
mechanical workshops of Pavia for the lead modules \cite{Silvietta}
and of Pisa for the copper ones. \\
 The geometry chosen for the lead and copper profile represented the
best compromise between the limitations given by the production technologies
and the need to maximize the sampling fraction and the sampling frequency
of the calorimeter. One of the most challenging aspects of building
this type of calorimeter is the problem of how to get very large numbers
of optical fibers embedded in a uniform way in the metal absorber
structure. 

\begin{figure}
\centerline{\includegraphics[width=0.95\textwidth]{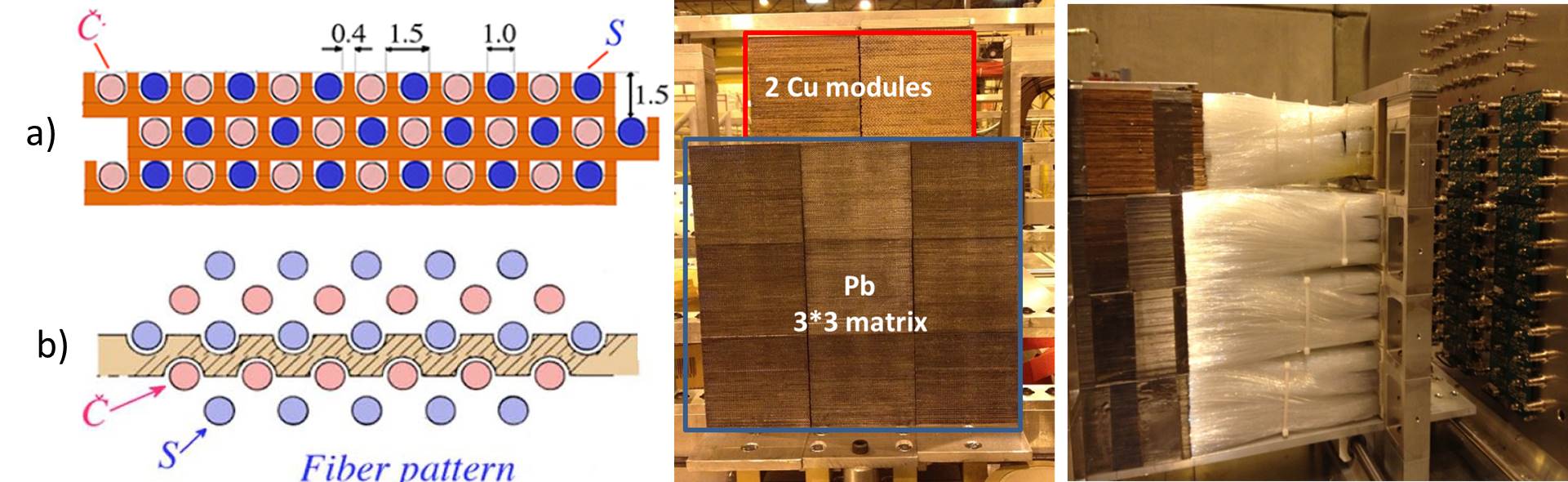}}
\caption{Left: fiber pattern used in our prototypes for the copper absorber
(a) and lead absorber (b). The two different colors represent the
two different kind of fibers. Right: transversal and lateral view
of the Pb-fiber and Cu-fiber RD52 calorimeter prototypes as installed
at the CERN test beam area. \label{RD52_fibers}}
\end{figure}

 In case of the lead absorber, a regular structure with equidistant
fibers was chosen (Fig \ref{RD52_fibers} b), while for the copper
absorber another solution was adopted: grooves were created only on
one side of each absorber plate, due to the difficulty to machine
both sides of the copper plates. In this case the Cherenkov and scintillating fibers were alternated in a single layer (Fig \ref{RD52_fibers})
a. The sampling fraction was calculated to be 4.5\% and 5\% in the
case of copper and lead geometry respectively. 
 
\subsection{Choice of fibers}
Before the beginning of the prototypes construction, we have chosen
the type of fibers more suitable for our application, in term of numerical
aperture and absorption length. A test bench was built at the INFN
Pisa in order to measure the fiber absorption length. A led and the
light produced by the Ru$^{106}$ $\beta$ decay (above the Cherenkov
threshold) were used. For the scintillation signal fibers Kuraray
doped SCSF-78 were chosen. Before the PMT each S channel was also
equipped with a yellow filter to eliminate effect of self absorption,
higher at shorter wavelengths. For the Cherenkov light, Mitsubishi
Polymethyl Methacrylate Resin (PMMA)\cite{PMMA} based fibers SK40 were chosen. Their attenuation length was measured to be $\sim$ 6m. 

 In order to farther increase the Cherenkov light yield, clear fibers
used in one of our copper prototypes have been aluminized on one side
with the spattering technique at Fermilab. 

\subsection{The Pb-fibre matrix}

\begin{figure}
\centerline{\includegraphics[width=0.95\textwidth]{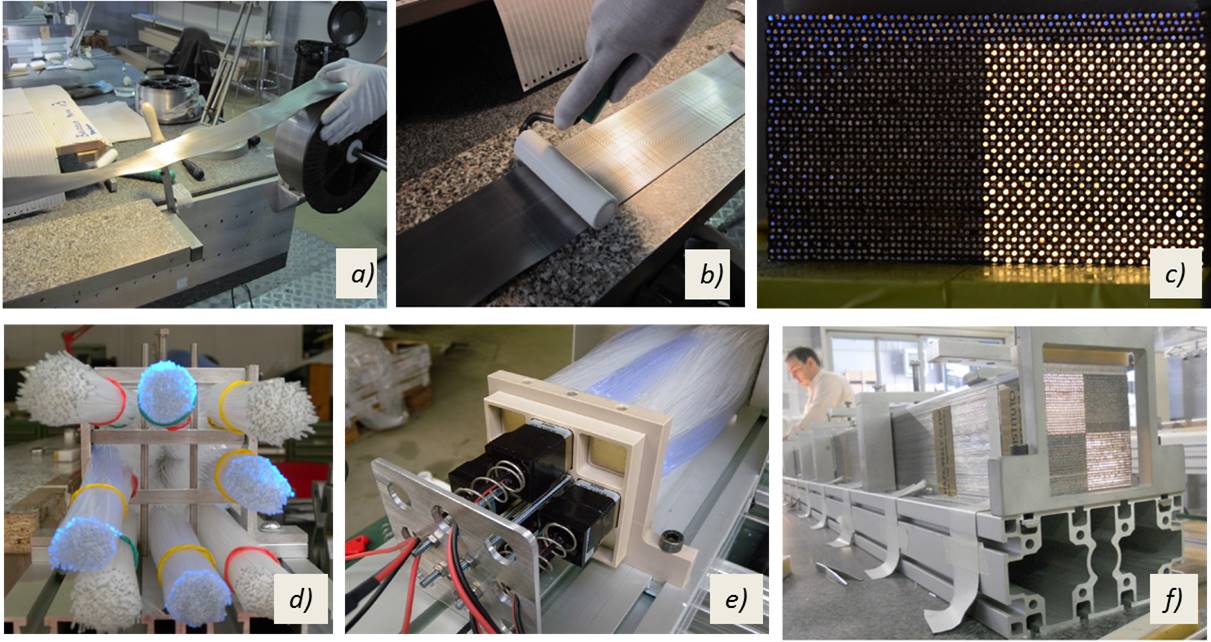}} \caption{Pictures taken during different phases of the assembly of lead/fibers
RD52 modules. \label{Pb}}
\end{figure}

The nine lead modules were arranged for the test beam in such a way
to form a 3x3 matrix as is shown in Fig \ref{RD52_fibers} right,
on top of them the two copper prototypes were lated. %Each module was created by stacking layers of lead with a proper geometry with layer of scintillating (XXX) and Cherenkov fibers (XXX). \\
For the construction of the nine lead modules, lead plates were produced
by an Italian company by the cold extrusion process; after few months
of development, the mechanical tolerances were measured to be suitable
for our needs: nominal thickness of 1 mm with a tolerance of 50 $\mu$m,
a fiber-to-fiber distance of 2 mm and a total width of 9.3 cm. Extruded
lead was shipped by the producer rolled up on a reel; during the assembly
phase each plate was cut at the nominal length (2.5 m), stretched
with a roller (Fig \ref{Pb} a and b) and placed on a worktable (Fig
\ref{Pb} e), where the stacking process occurred. Each layer of fibers
was created, by laying carefully 46 fibers of a given type in each
of the lead grooves. The stacking continued, alternating layers of
Cherenkov and scintillating fibers, until the height of the module
reaches 9.3 cm.

 For each module, eight fiber bundles (4 scintillating and 4 Cherenkov)
depart from the rear end and are glued to a support and are milled
before being connected to a PMT (Fig \ref{Pb} c and d). 

\subsection{The Cu-fibre modules}
\begin{figure}
\centerline{\includegraphics[width=0.75\textwidth]{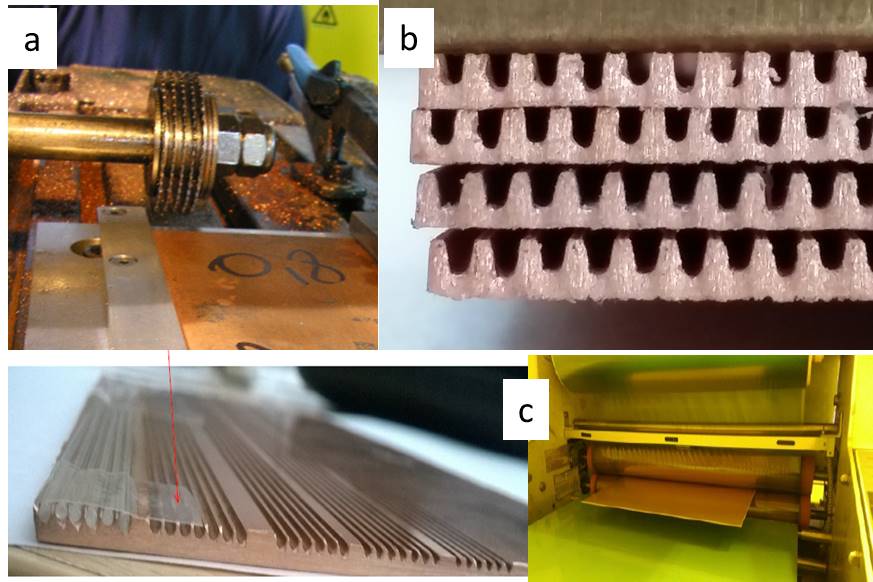}} \caption{Different methods for the fabrication of the RD52 copper absorbers:
a) The saw scraper method used at the INFN Pisa for building the first
two copper modules prototypes. b) First results of grooves created
in the copper planes with a water jet (Iowa University). c) Grooves
in the copper plane created with chemical milling (CERN) \label{Cu_grooves}}
. 
\end{figure}

For the copper prototypes construction has turned out that copper
itself is a particular difficult material to work with such high density
aspect ratio. We have tried many different ways (e.g. rolling, extrusion),
but so far only machining grooves in thin copper plates has provided
the desired quality. This is why the design was slightly modified
in order to etch grooves only on one side of the copper plates (see
Fig \ref{RD52_fibers} a). For the first two prototypes, copper planes
were grooved with a rotating saw (water cooled) as is shown in Fig
\ref{Cu_grooves} a. The rotating saw was fabricated specifically
for this application, with four parallel lames on a common rotating
tool. Once the copper planes were prepared (50 cm long), the assembly
procedure was similar to the one used for the lead module, with some
differences, like the fact that copper plates were laid one after
the other in order to reach the 2.5 m of total longitudinal length,
and that between one layer and the other some glue was used and pressed
during the night. 

\subsection{Industrialization of the construction process}
Copper is a better material than lead for the dual readout calorimetry because of the much better electromagnetic resolution. 
However, in order to be able to produce in the next future a real
scale, full containment copper calorimeter with such high sampling
frequency as the one we tested so far, we need to find a less time
consuming and an industrial-compatible technique for making grooves
in copper.

 The method used with rotating saws is time consuming if is needed
to be applied in large scale. We have then tried other ways and from these studies the more promising
ones seem to be chemical milling and water grooving. After having
started the rough grooving process with one of the two methods one
could finish the grooving process with rolling to better define the
precision of the grooves. First trials of chemical milling were done
at CERN using the classic photolitography technique (Fig \ref{Cu_grooves}
c) and tuning the width of the mask and the exposure time to the chemical
bath. First trials of water grooving were done at the Iowa State University
(Fig \ref{Cu_grooves} b) tuning the water pressure and the jet speed.

Finally, toward a fully projective fiber calorimeter structure, the possible geometry has been studied years ago by the RD1 collaboration
\cite{RD1} and this geometry could be still applied for the dual readout fiber calorimeter. 

\subsection{Future directions}
 In order to study the possibility to use  a dual readout, not longitudinally segmented calorimeter, for a future collider experiment, the RD52 collaboration is also equipping one of the copper modules with Silicon photomultipliers readout.  This readout way offers the possibility to eliminate the forests of optical fibers that stick out at the rear end and also to chose a trasnversal segmentation at a fiber level, if it is needed. With PMT readout, fiber bunches (Fig\ref{Pb}) occupy precious space and act as antennes for particles that comes from other sources than the showers developing in the calorimeters. Tests of this new readout is part of the experimental program foreseen for the coming fall.
 
\section{Dual Readout Calorimetry with Crystals}
Using high-Z crystals as dual-readout calorimeters offers potential
benefits, since one could then in principle eliminate or greatly reduce
two remaining sources of fluctuations that dominated the hadronic
resolution of the fiber calorimeter: sampling fluctuations and fluctuations
in the Cherenkov light yield. In case of using homogeneous media,
the light signals generated in the crystals into scintillation (S)
and Cherenkov (C) components need to be decomposed. In recent years
we have developed four different methods to separate these two components,
exploiting their different properties: 
\begin{itemize}
\item differences in the angular distribution of the emitted light, 
\item differences in the spectral characteristics, 
\item differences in the time structure of the signals, 
\item light polarization in case of Cherenkov 
\end{itemize}
These methods were experimentally investigated and optimized -for
three different types of crystals: bismuth germanate (Bi$_{4}$Ge$_{3}$O$_{12}$,
BGO) \cite{2008}, bismuth silicate (Bi$_{4}$Si$_{3}$O$_{12}$,
BSO) \cite{NIM_2011_BSO}, \cite{NIM_2011_pol}, and lead
tungstate (PbWO$_{4}$) \cite{2008}. The latter
crystal was also doped with small amounts of impurities (Molybdenum
and Praseodymium) to further improve its dual-readout characteristics
\cite{NIM_2010_xtals}. However, of these four methods only
the differences in the spectrum and in time structure are easily applicable
in hermetic detectors needed for 4$\pi$ experiment at particle colliders.

\subsection{Undoped PWO crystals}
\begin{figure}
\centerline{\includegraphics[width=0.75\textwidth]{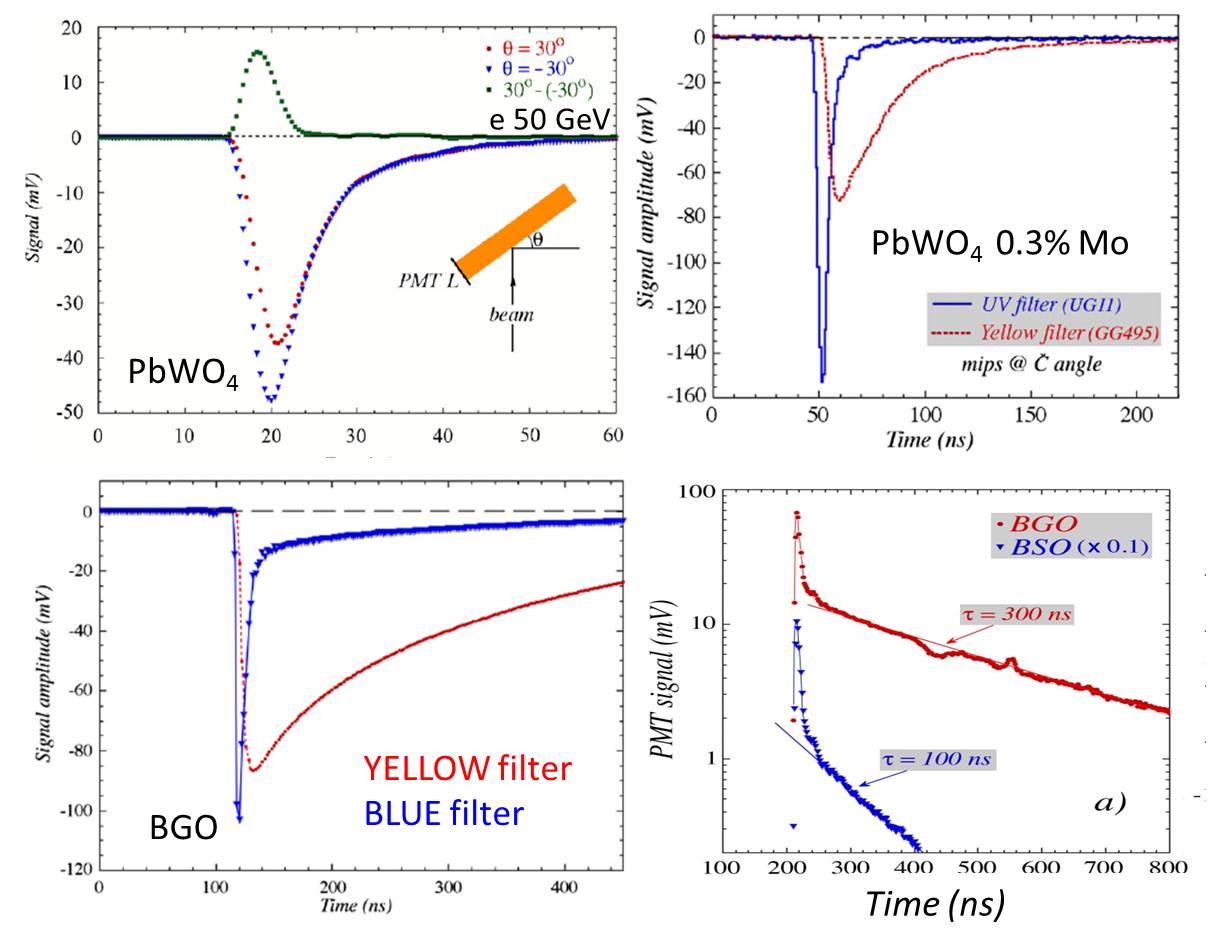}}
\caption{Results from the single crystals measurements. a) Time structure from
a PMT for two orientations of the PbWO$_{4}$ crystal with respect
to the incoming beam; the difference between the two distributions
is an indication of the presence of Cherenkov light in the forward
direction (green squares), b) Mo-doped PbWO$_{4}$ pulse shapes for
the crystal side equipped with yellow filter and the UV one. c) BGO
pulse shapes for yellow and blue filter. d) comparison of scintillation
amount and time scale for BGO and BSO crystals \label{Xtals}}
\end{figure}

 The first tests that proved the feasibility of separation of Cherenkov
from Scintillation light in homogeneous media were done on a single
PbWO$_{4}$ crystal \cite{NIM_2007_PWO},   \cite{2008},
and exploited the directionality of Cherenkov light. Results are shown
in Fig \ref{Xtals} a, where one can see the average time structure
from the crystal oriented at +30 and -30 degrees with respect to the
incoming electron beam. The difference of the two signals is an indication
of the Cherenkov component, that makes the leading edge of the forward
signal much steeper than the backward one.

 After these promising first studies, an intense R\&D program of Dual
Readout calorimetry with crystals started. In order to exploit the difference in spectral emission of scintillation (that has a characteristic emission spectrum, depending from the type of crystal) from Cherenkov light (wih a 1/$\lambda^{2}$ behavior), four optical transmission filters were used in our studies, all were
3 mm thick and made of glass; each of them had 90\% of transmission
for different wavelengths: UG11: $\lambda<400$ nm , UG330: $\lambda<410$
nm, UG5: $\lambda<460$ nm, GG495: $\lambda>495$ nm. For the isolation
of the Cherenkov component, a cut towards shorter wavelengths ended
up with the effect to have less contamination of scintillation light
but, on the other end, Cherenkov signals obtained with this filter
were rather small and strongly depended on the distance the light
had to travel to the PMT. Different combination of filters were used
in order to find the right compromise. 
 
\subsection{BGO crystals}
The importance of optical transmission filters for our purpose was
first demonstrated with BGO crystals \cite{2008}
and then successfully applied to Mo-doped PbWO$_{4}$ ones \cite{NIM_2010_xtals}.
Fig. \ref{Xtals} b and c shows the typical signal shapes for events
in which beam particles traversed a single crystal placed perpendicular
to the beam line. One side of the crystal was equipped with a ``yellow''
(GG495) transmission filter, the other side with a UV (UG11) one.
The UV filter absorbed more than 99\% of the scintillation light,
while a large fraction of the Cherenkov light was transmitted. As
a result, the Cherenkov component of the light produced by the crystal
became clearly visible, in the form of a prompt peak superimposed
on the remnants of the scintillation component, which has a longer
decay time (25 ns for Mo-PbWO$_{4}$ and 300 ns for BGO). 

\subsection{Doped PWO crystals}
 In \cite{NIM_2010_xtals} one can read about the studies we
did in order to optimize PbWO$_{4}$ crystals for dual readout applications,
introducing small amounts of molybdenum and praseodymium doping. The
best solution found was 0.3\% Mo-doping, which had two important effects
for the separation of Cherenkov from scintillation light: the shift
of the emission spectrum towards longer wavelengths, making possible
to use of optical filters, and the increase of the decay time of the
scintillation process from $\sim10$~ns to $\sim25$~ns, making 
easier the separation of the prompt Cherenkov component from the delayed
scintillation one. In Fig \ref{Xtals} b one can see that for the
Mo-doped PbWO$_{4}$ crystals, we could achieve an almost complete
separation of the two optical components.

\subsection{Time structure of the light signal}
 In order to exploit the different time structures between prompt
Cherenkov light and scintillation, the pulse shape of the signals
produced in these crystals irradiated with electron beams were recorded
by means of an high sampling frequency oscilloscope (4 ch) or by 32
channels digitizer based on the DRS-IV chip, with a time resolution
up to 200 ps.

\begin{figure}
\centerline{\includegraphics[width=0.75\textwidth]{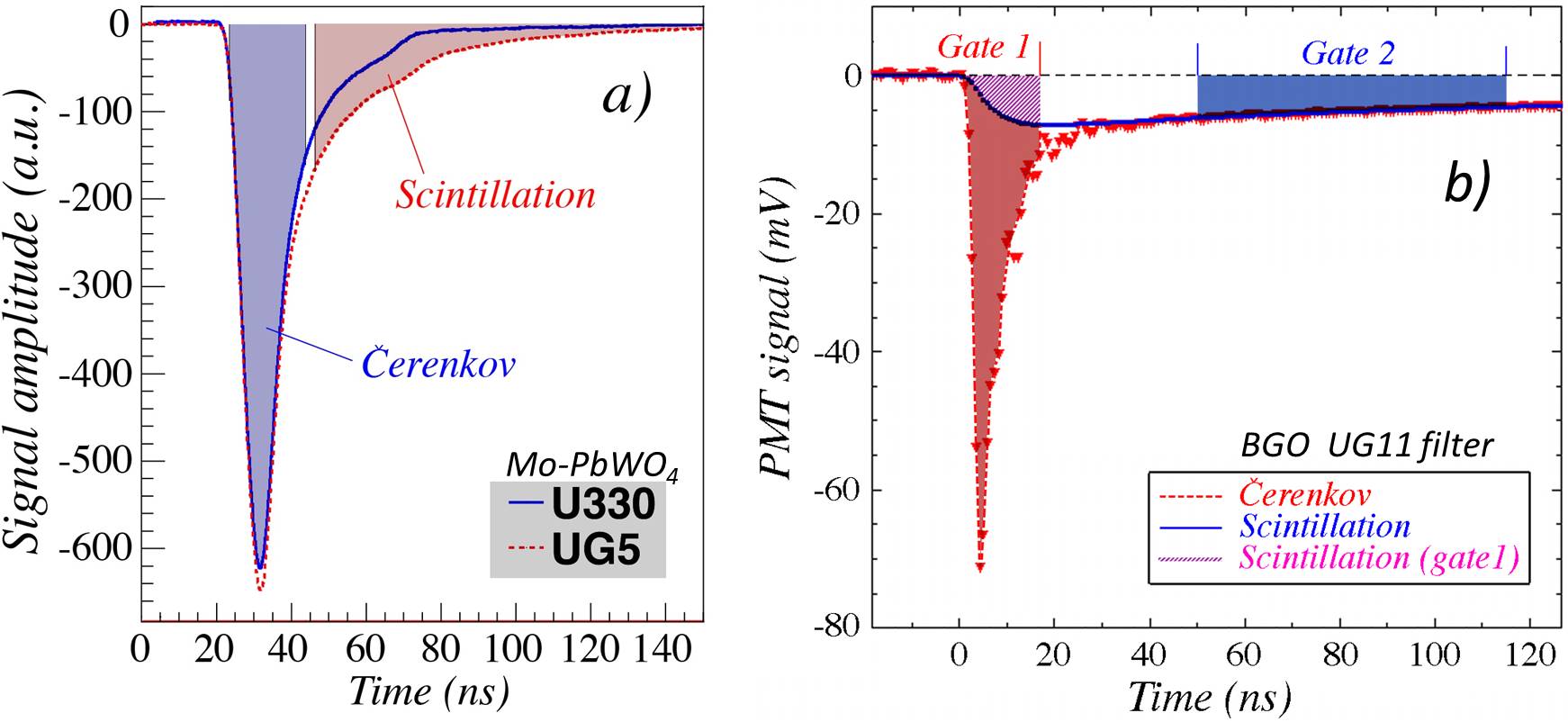}} \caption{ The time structure of a typical shower signal measured in the Mo-PbWO$_{4}$
equipped with two optical filters (a) and of a BGO crystal equipped
with an UG11 filter (b) .In order to measure the relative contributions
of scintillation and Cherenkov light, the time spectrum is integrated
in two different gates. \label{Time_gates}}
\end{figure}

 The time structure information of the signals was used to determine
their scintillation and Cherenkov components, integrating, event by
event, the pulse shape in two different gates, as it can be seen in
Fig \ref{Time_gates}. The Cherenkov signal is defined to be the result
of the integral of the pulse in the first gate, while the scintillation
signal is defined by the integral on the second and more delayed gate.
These integrated charges were then converted into deposited energy
using appropriate calibration constants (see later for more details).

Using the time structure of the signal has the advantage that the
C and S components are extracted by the same readout channel, while
for the optical filter case two different channels, equipped with
different filters are needed. However, as it can be seen from Fig
\ref{Time_gates}, not always the use of time structure from only
one readout channel is easy. The result will depend from how much
scintillation light is passing from the optical filter bandwidth and
from how far in time is the scintillation component from the prompt
Cherenkov one. In the case of BGO crystal (Fig \ref{Time_gates} b),
because of a longer scintillation decay time, one can see that the
definition of two well separated time gates is more easy than in the
other case. A much higher scintillation light yield is also helpful
in the BGO case because the signal filtered by the UV filter is much
more ``contaminated'' by scintillation photoelectrons, that give
the characteristic long tail to the time distribution.
 
 \subsection{Comparison of BGO and BSO crystals}
Another type of crystal that has also been studied for dual readout
applications \cite{NIM_2011_BSO} was the bismuth silicate (BSO),
which has the same crystal structure as BGO (Bi$_{4}$Si$_{3}$O$_{12}$),
with silicon atoms replacing the germanium ones.

 We have performed a systematic comparison of the BGO and BSO crystals
relevant properties and we found that the BSO crystal could be very
promising if one would like to invest more studies in the optimization
of crystals for dual readout calorimetry.

 First the purity of the Cherenkov signals that can be obtained with
UV filters was studied. We found the contamination of scintillation
light in the Cherenkov signals to be smaller by about a factor of
two in the BSO crystal. Then we studied the number of Cherenkov photoelectrons
detected per unit of deposited energy; we measured this yield to be
about a factor two to three larger for the BSO crystal. The light
attenuation length of Cherenkov light has been found to be approximately
the same as the BGO. The cheaper price of BSO with respect of BGO
could also be an advantage. In Fig \ref{Xtals} da comparison between
the emission spectrum of BGO ($\sim$ 300 ns ) and the BSO ($\sim$
100 ns) it shown. The slightly faster signal of BSO could be an advantage
for high energy applicatons, the time scale of hundred of nanoseconds
is still ideal for using informations coming only from one readout
channel and, as in the case of BGO, using the integral of pulse shape
in two well defined time windows.

\subsubsection{Electron showers in dual-readout crystal calorimeters}

The application of the dual readout technique to an hybrid system
made by a crystal matrix (BGO or Mo-doped PbWO$_{4}$) as electromagnetic
section and the original DREAM fiber calorimeter as the hadronic one
was evaluated \cite{NIM_BGO_DREAM}.

 Thanks to the experience gained with previous single crystal studies
on the way to separate the Cherenkov from the scintillation light
components, we have also applied to the crystal matrices the same
methods: the optical filters and time structure to select the desired
type of light.

 The BGO matrix was made by 100 BGO crystals, 24 cm long, from a projective
segment of the L3 experiment. The segment was placed perpendicular
to the beamline (25 X$_{0}$ deep), as shown in Fig \ref{BGO_DREAM}
b, in front of the DREAM calorimeter The readout was made in different
ways, the most promising results were obtained with 16 square PMTs
(Photonis XP3392B) arranged in a way such that each PMT collected
light produced by clusters of at least 9 adjacent crystals. Each PMT
was equipped with an UG11 optical filter.

 The PbWO$_{4}$ matrix (Fig \ref{BGO_DREAM} a) consisted of seven
custom made 0.3 \% Mo-doped crystals, 20 cm long, arranged in a matrix
and placed in the beam line with the beam entering in the central
crystal (22.5 X$_{0}$ longitudinal dimensions). Each crystal was
equipped with one PMT on each side (Hamamatsu R8900-100); both faces
of the matrix were covered with a large optical transmission filter.
Several filter combinations were used during our tests. A calibration 
procedure was done for both matrices using electron beams.

\subsubsection{Calibration}
 For the PbWO$_{4}$ one a narrow beam was steered into the center
of each of the seven crystals constituting the matrix. According to
GEANT4-based Monte Carlo simulations 93\% of the energy was deposited
in the entire matrix. After the signals from the crystal were disentangled
into Cherenkov and scintillation components, the integrated charge
in each of these components was determined, equalization constants
were found and applied before summing he contribution of each crystal
to get the total inter-gate charge. The average value of that total
charge for beam particles traversing the center of the matrix was
equated to 93\% of the beam energy, and this yielded the conversion
factor between the normalized integrated charge and the deposited
energy. 

 In the case of the BGO matrix, calibration constants had to be assigned
to each of the four PMTs that read out the four longitudinal segments
of the matrix into which the showers developed. This calibration procedure
was carried out in two steps: first, the gains of all 16 PMTs were
equalized, by means of an LED signal with an amplitude comparable
to that of a typical electron shower signal. In the second step, 100
GeV electrons were sent into each of the four columns and the HV values
of the four PMTs in the hit column were varied, in an iterative procedure,
until the energy resolution for the summed signals reached a minimum
value. Because of the size of the BGO matrix, size leakage was considered
negligible, and we assumed that the integrated charge collected by
the 16 PMTs was a good measure of the deposited energy. On that basis,
the integrated charge measured in each individual PMT contributing
to the signal could be converted into GeV as well.

\begin{figure}
\centerline{\includegraphics[width=0.75\textwidth]{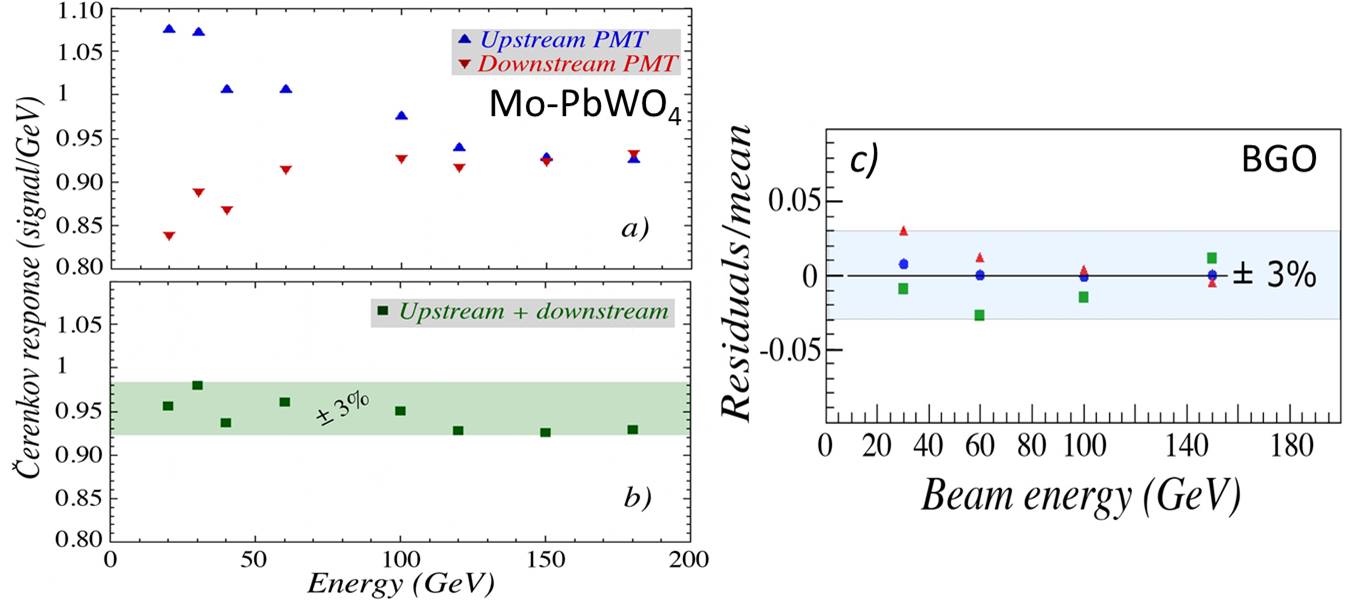}} \caption{ Signal linearity of electron detected with the crystals matrices.
Shown is the response as a function of the electron energy for the
C signal in a Mo-PbWO$_{4}$ matrix equipped with U330 filters on
both sides (a and b); S, C and $\Sigma$ signals in case of BGO matrix
equipped with UG11 filters (c). The signals measured at both ends
are both shown separately (a) and added together (b). \label{Lin}
\cite{NIM_2012_xtals}}
\end{figure}

\subsubsection{Linearity of the response}

We have investigated the linearity of the observed signals in both
crystal matrices and results are shown in Fig \ref{Lin}. In the case
of BGO the calorimeter is linear within a 3\% for both the C and S
components (Fig \ref{Lin} c). In the case of Cherenkov signal in
PbWO$_{4}$, the linearity is restored only if the signal extracted
from the PMTs placed on both sides of the crystals are summed (Fig
\ref{Lin} b). This effect is the result of the strong attenuation
of the UV light in these kind of crystals, hence as the energy increases,
the shower maximum is closer to the downstream PMT and its response
is higher, vice versa for the downstream one, as can be seen in Fig
\ref{Lin} a. In order to obtain a linear Cherenkov response we have
equipped both sides of the matrix with a low pass wavelength filter
(U330). The main disadvantage of this readout geometry is that the
U330 filters transmit almost no scintillation light, as can be seen
from Fig \ref{Time_gates} a. In an alternative setup, we therefore
replaced the downstream U330 filter by a UG5 one, which also transmits
light with wavelengths in the region around 500 nm, where scintillation
dominates. This led to a usable scintillation signal.

\begin{figure}
\centerline{\includegraphics[width=0.95\textwidth]{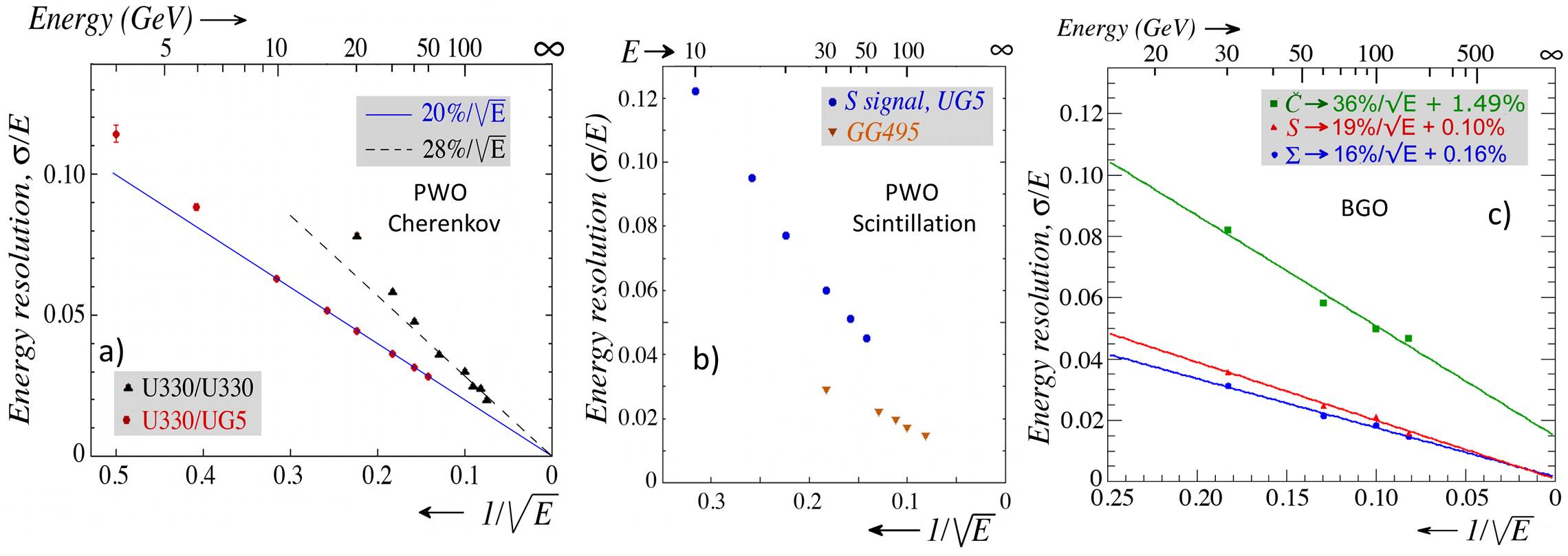}} \caption{Energy resolution for electrons in the crystal matrix: MO-doped PbWO$_{4}$
in a and b figures and BGO in c.a): resolution measurured Cherenkov
signals, derived from UV-filtered light detected at both ends of the
crystal matrix. b) scintillation signal measured with two different
filters. c: results for the total charge collected by the PMTs ($\Sigma$)
and for Cherenkov (C) and scintillation (S) components of the signal.
Results from the fit of experimental points are also shown. \label{EresoXtals}
\cite{NIM_2012_xtals}} 
\end{figure}

\subsubsection{Energy resolution}
The energy resolution obtained for the Cherenkov channel is shown
as a function of the electron energy in Fig. \ref{EresoXtals} a for
these two sets of experimental data. The stochastic fluctuations that
dominate the measured energy resolutions were found to contribute
20\%/$\sqrt{E}$ and 28\%/$\sqrt{E}$ for these two filter configurations.
No evidence was found for an energy independent contribution to the
energy resolution. The resolutions measured at the low-energy end
of each data set deviate from the straight lines. These deviations
are consistent with the contribution of the signal baseline fluctuations
to the measured energy resolution. The energy resolution is strongly
dominated by fluctuations in the Cherenkov light yeald, this can be
seen by the fact that the Cherenkov energy resolution improves if
UG5 filter, that transmit a larger fraction of Cherenkov light, is
used.

 Even though the UG5 filter led to usable scintillation signals, the
energy resolution for electron showers measured in the scintillation
channel (as an integral of the tail of the signal) was somewhat worse
than in the Cherenkov channel, and about a factor of two worse compared
to resolutions measured with the yellow (GG495) filter (Fig \ref{EresoXtals}
b). This is of course due to the very small fraction of the scintillation
light that was detected in this setup. From the sloped of the lines
drown in Fig \ref{EresoXtals} a the Cherenkov light yield was estimated
to be 25 photoelectrons per GeV in case of UG5 filter and 13 for the
U330\footnote{ From this small amount of Cherenkov photoelectrons one could see
that it could be a limiting factor for the dual readout calorimetry.
Fluctuations in this small number could affect calorimeter performances.
We will see in the next chapter that the RD52 collaboration found
higher values of Cherenkov phooelectrons in a sampling dual readout
calorimeter, that is nowadays the priority of the collaboration.}.

 In case of BGO matrix the energy resolution measured with electrons
in the range 10 - 150 GeV is shown in Fig \ref{EresoXtals} c: 
\begin{equation}
\frac{\sigma E_{C}}{E}=\frac{36\%}{\sqrt{E}}+1.5\%\quad\frac{\sigma E_{S}}{E}=\frac{19\%}{\sqrt{E}}+0.1\quad\frac{\sigma E_{\Sigma}}{E}=\frac{16\%}{\sqrt{E}}+0.16\%
\end{equation}

The resolution obtained for the scintillation component, as well as
that for the total collected charge ($\Sigma$) i.e. the integral
over the entire waveform, are well described by $E^{=\frac{1}{2}}$
scaling, while the energy resolution measured for the Cherenkov component
exhibits a deviation, with a constant term b $\sim$ 1.5\%. The resolution
for this component is affected by significant non-Poissonian fluctuations.
This could be due to the fact that in order to use crystals for dual
readout calorimetry, we had to apply cuts in the wavelength and in
the amount of emitted light; in particular for the C component we
selected a very small fraction of of the total light produced in the
crystals, in a wavelength range in which that light is strongly attenuated.

In the case of the BGO matrix we have also studied the performance
of the whole calorimeter with pion beams \cite{NIM_BGO_DREAM}.
The information collected with the two sections are combined to obtain
an overall evaluation of the electromagnetic fraction of the shower.
In order to perform such a measurement we selected only those events
in which pions start showering in the crystal section.
 
\begin{figure}
\centerline{\includegraphics[width=0.75\textwidth]{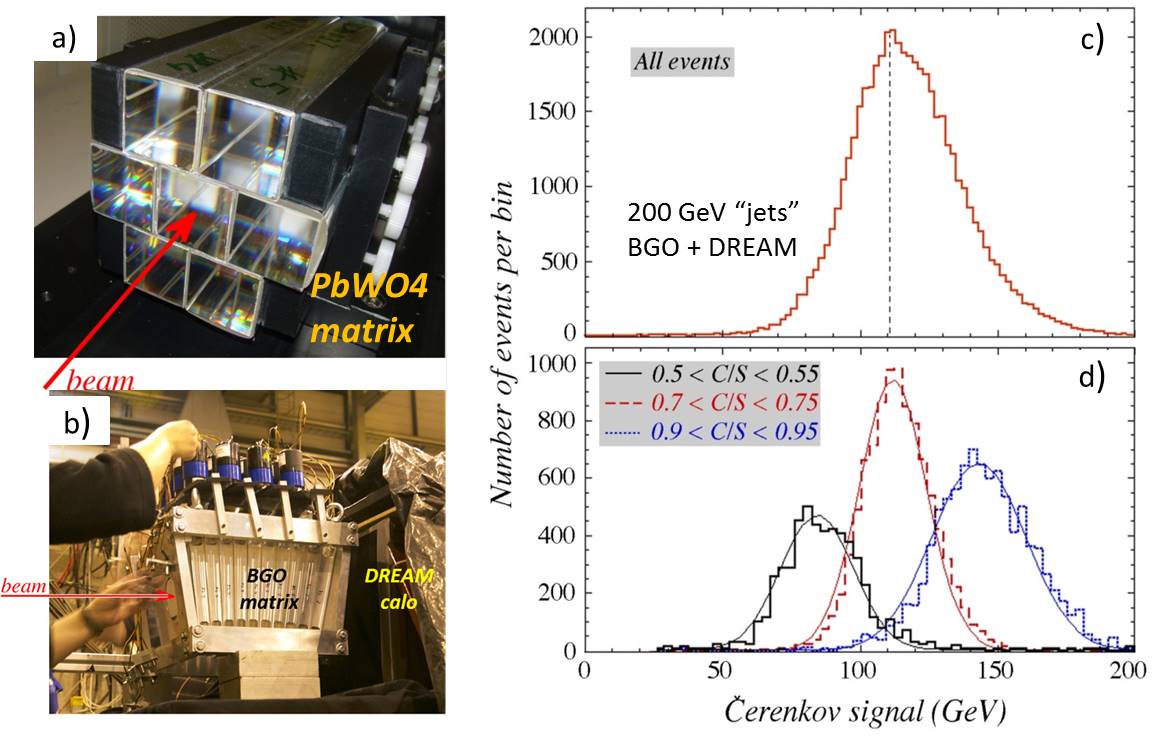}} \caption{ Picture of the PbWO$_{4}$ matrix (a) and of the hybrid BGO matrix
and fiber DREAM calorimeter (b). Cherenkov signal distribution for
200GeV jet events detected in the BGO plus fiber calorimeter system
(c), distributions for subsets of events selected on the basis of
the ratio of the total Cherenkov and scintillation signals (d). \label{BGO_DREAM} }
\end{figure}

In Fig \ref{BGO_DREAM} c the distribution of the total Cherenkov
signal of the hybrid system BGO + DREAM fiber calorimeter is shown.
This signal is broad, asymmetric and centered around a value of only
110 GeV, whereas the total jet energy was 200 GeV. In Fig \ref{BGO_DREAM}
d three different subsets of events, selected on the basis of the
measured C/S signal ratio are shown. These three distributions are
narrower, well described by Gaussian fits and centered at a value
that increases roughly proportionally with the C/S value of the selected
event sample. This is precisely what was observed for the fiber calorimeter
in stand-alone mode, and what allowed to eliminate the effects of
fluctuations in the electromagnetic shower component in that calorimeter.

\subsection{Future prospects for dual-readout crystal calorimeters}
In conclusion to our R\&D on crystals for dual readout calorimetry,
we can say that our interest in studying high-Z scintillating crystals
for the purpose of dual-readout calorimetry derived from the potential
reduction of the contribution of stochastic fluctuations to the energy
resolution of such calorimeters. Our goal in further developing this
experimental technique is to reduce the contribution of stochastic
fluctuations to the point where these are comparable to the irreducible
effects of fluctuations in visible energy. Crystals were believed
to offer good opportunities in this respect. However, as the results
of this study show, things are not so easy.

For the application of crystals to the dual readout method, the detection
of Cherenkov light is the crucial issue and particular care was taken
in the precision with which the calorimeter performance can be measured
using this signal component. Extracting sufficiently pure Cherenkov
signals from these scintillating crystals with the use of optical
filters implies a rather severe restriction to short wavelengths.
As a consequence, a large fraction of the potentially available Cherenkov
photons needs to be sacrificed, but also, the light that does contribute
to the Cherenkov signals is strongly attenuated, because of the absorption
characteristics of the crystals. As a result, the remaining light
yield is such that fluctuations in the detected numbers of photoelectrons
become a significant contribution to the em energy resolution. This
is an important difference with experiments in which the unfiltered
light of such crystals is used for the electromagnetic calorimeter
signals. 

 The conclusion reached after after a long and in-depth study of crystal
performances for dual readout calorimetry is that no such significant
improvements in term of Cherenkov light yield seem to be offered by
crystals in combination with filters in dual-readout calorimeters.
The RD52 collaboration decided therefore to focus on the fiber option.

\section{Summary and Future Plans}

We have proved that the Dual-REAdout Method calorimeter could achieve
the high-quality energy measurement for both electromagnetic particles
and hadrons. Even, GEANT4 simulation anticipates that the dual-readout
calorimeter can have $30\%/\sqrt{E}$ for the stochastic term and
very small constant term. This performance will satisfy the important
requirement of the future lepton collider, which is the separation
of hadronically decaying W/Z bosons. The RD52 collaboration is make
effort to build a large size of calorimeter with Cu and will try to
prove that it would achieve the excellent energy measurement of hadrons
and jets in the future.

The more excellent detector performance, the better physics results
we have. We expect that the Dual-REAdout Method calorimeter will be
able to achieve the high-precision energy measurement for all fundamental
particles and open a new era of experimental particle physics, just
as the high purity Germanium crystal detector achieved the excellent
energy resolution in the nuclear spectroscopy and showed very detail
energy levels of nuclei.

\section*{Acknowledgments}

We thank CERN for making good particle beams available to our experiments
in the H8 beam. This study was carried out with financial support
from the United States Department of Energy, under contract DE-FG02-12ER41783,
by Italy's Istituto Nazionale di Fisica Nucleare and Ministero dell'Istruzione,
dell'Universit\'a e della Ricerca, and by the Basic Science Research
Program of the National Research Foundation of Korea (NRF), funded
by the Ministry of Science, ICT \& Future Planning under contract
2015R1C1A1A02036477. In addition, we thank Korea University for the
support received by their researchers who contributed to this project.

\end{document}